\documentclass[a4paper,11pt]{article}
\pdfoutput=1 

\usepackage{jcappub} 

\usepackage[T1]{fontenc} 
\usepackage{amsmath, amssymb, bm, graphics, color,mathrsfs,hyperref, slashed}
\usepackage{subfig,float}

\newcommand{\be}{\begin{equation}}
\newcommand{\ee}{\end{equation}}
\newcommand{\ben}{\begin{eqnarray}}
\newcommand{\een}{\end{eqnarray}}

\newcommand{\la}{{\lambda}}

\newcommand{\cO}{{\cal O}}

\newcommand{\cL}{{\cal L}}

\newcommand{\na}{\nabla}

\newcommand{\tal}{{\tilde \alpha}}

\newcommand{\tchi}{\tilde \chi}
\newcommand{\trho}{\tilde \rho}

\newcommand{\hM}{{\hat M}}

\newcommand{\hht}{{\hat t}}
\newcommand{\hr}{{\hat r}}
\newcommand{\ha}{{\hat a}}
\newcommand{\hSigma}{{\hat \Sigma}}
\newcommand{\hDelta}{{\hat \Delta}}
\newcommand{\hmu}{{\hat \mu}}
\newcommand{\hrho}{{\hat \rho}}

\newcommand{\htheta}{\hat \theta}

\newcommand{\ttheta}{{\tilde \theta}}

\newcommand{\ep}{\epsilon}

\newcommand{\talpha}{\tilde \alpha}
\newcommand{\tbeta}{\tilde \beta}
\newcommand{\tla}{\tilde \lambda}

\newcommand{\ga}{\gamma}

\title{Dark matter cosmic string in the gravitational field of a black hole}

\author[a,1]{{\L}ukasz Nakonieczny\note{Corresponding author.}}
\author[a]{, Anna Nakonieczna}
\affiliation[a]{Institute of Theoretical Physics, Faculty of Physics, University of Warsaw \protect \\
ul. Pasteura 5, 02-093 Warszawa, Poland }
\author[b]{and Marek Rogatko}
\affiliation[b]{Institute of Physics, Maria Curie-Sk{\l}odowska University \protect \\
pl. Marii Curie-Sk{\l}odowskiej 1, 20-031 Lublin, Poland}

\emailAdd{Lukasz.Nakonieczny@fuw.edu.pl}
\emailAdd{Anna.Nakonieczna@fuw.edu.pl}
\emailAdd{rogat@kft.umcs.lublin.pl}

\abstract{We examined analytically and proposed a numerical model of an Abelian Higgs {\it dark matter} vortex in the spacetime of a stationary axisymmetric Kerr black hole.
In analytical calculations the {\it dark matter} sector was modeled by an addition of a $U(1)$-gauge field coupled to the visible sector. The backreaction analysis revealed that 
the impact of the {\it dark vortex} presence is far more complicated than causing only a deficit angle. The vortex causes an ergosphere shift and the event horizon velocity is also influenced by its presence.
These phenomena are more significant than in the case of a visible vortex sector. The area of the event horizon of a black hole is diminished and this decline is larger in comparison to the Kerr black hole with an Abelian Higgs vortex case. 
After analyzing the gravitational properties for the general setup, we focused on the subset of models that are motivated by particle physics. We retained the Abelian Higgs model as a description of the {\it dark matter} sector (this sector contained a heavy {\it dark photon} and an additional complex scalar)
and added a real scalar representing the real component of the Higgs doublet in the unitary gauge, as well as an additional $U(1)$-gauge field representing an ordinary electromagnetic field. Moreover, we considered 
two coupling channels between the visible and {\it dark sectors}, which were the kinetic mixing between the gauge fields and a quartic coupling between the scalar fields. 
After solving the equations of motion for the matter fields numerically we analyzed properties of the cosmic string in the {\it dark matter} sector and its influence on the visible sector fields that are directly coupled to it. We found out that the presence of the cosmic string induced spatial variation in the vacuum expectation value 
of the Higgs field and a nonzero electromagnetic field around the black hole.
}


\begin{document} 
\maketitle

\section{Introduction}

The idea of topological defects like monopoles, domain walls or cosmic strings as probes of the early Universe reaches back to the second half of the previous century. 
Among them cosmic strings are especially interesting from the perspective of both cosmology and particle physics. Although not observed yet, their existence may serve as a source of some exotic astrophysical phenomena like high energy gamma rays~\cite{bha89,Long_Hyde_Vachaspati_2014,Long_Vachaspati_2014,mot15}
or seeds for formation of supermassive black holes at high redshifts~\cite{bra15}. 
Other cosmological phenomena in which their presence in the form of a network may have left partial imprints are large scale structure formation and the thermal history of the Universe~\cite{vil94,gor11}.

With the precision measurements of parameters of the Standard Model of particle physics (SM) the possibility of existence of astrophysically observable
cosmic strings gets severely constrained. On the other hand, despite its tremendous success, the Standard Model is unable to explain all measurable astrophysical data,
namely it lacks the {\it dark matter} candidate. This can be cured by extending it by another sector describing the aforementioned type of matter. There are various 
approaches to this problem from very complicated extensions such as supersymmetric models~\cite{ter06} to the minimalistic ones like adding a single additional 
scalar representing the {\it dark matter} candidate~\cite{guo10}. Amongst these models many contain, beside scalar or fermionic candidates, also additional gauge fields. 
The possibility of existence of such a gauge field in the dark sector has recently stirred a large interest among the particle physics community. 
In the case when this new field is Abelian its particle is often called a dark photon or $Z^\prime$ and its existence and properties are intensively investigated 
from the theoretical point of view~\cite{bae13,cha14,bae15,arg17} and from the perspective of its detection in ongoing and planned experiments~\cite{PhysRevLett.113.201801,PhysRevD.92.115017}. 
In this context, cosmic strings that may form in such extended models may open a new window to the dark sector, and therefore investigations of their properties regain importance for particle physics, astrophysics and cosmology. 
On the other hand, topological defects in black hole spacetimes are interesting regarding the {\it no-hair} conjecture or its mathematical formulation known as the
uniqueness theorem for black objects. The theorem in four-dimensional gravity states that the only static electrovacuum black hole spacetime is given by the 
Reissner-Nordstr\"om solution, while the stationary axisymmetric one is described by the Kerr-Newman black hole line element~\cite{book}. The generalizations of the aforementioned theorem  to higher dimensional spacetimes were proposed in~\cite{gib02,rog03,rog06} for static black objects. For stationary axisymmetric cases the attempts to find the uniqueness theorem were presented in~\cite{mor04,mor08,rog04a,rog08,hol08,hol08a}. In five dimensions one finds the black ring solution with all characteristics the same as for a five-dimensional Kerr black hole, but with a different topology of the event horizon. The enhancement of the {\it no-hair} conjecture to comprise the fields with a nontrivial topology allows to check whether topological defects may authorize hair on black holes.
Not only cosmic strings, but also domain walls and their influence on black objects were widely studied in the literature of the subject~\cite{mor00,rog01,mod03,mod04,mod06,rog04,fla05,fla06a,czi09,czi10,czi11,gal14}.

The first theoretical description of gravitational properties of cosmic strings was proposed in~\cite{vil81,DESER1984220,Gott:1984ef}, where a metric of a cosmic string was derived
under the assumption that matter distribution is given by a Dirac delta like source. 
After obtaining a satisfactory description of the gravitational properties of cosmic strings the next step was to investigate their interactions with other astrophysical objects.
First attempts in this direction were presented in~\cite{ary86}, where the metric describing a Schwarzschild black hole threaded by a cosmic string
was provided. An extension of the problem to the case of an Abelian Higgs vortex on the Euclidean Einstein~\cite{dow92} and Euclidean dilaton black hole~\cite{mod98} systems was performed.
Next, numerical and analytical studies revealed that the Abelian Higgs vortex could act as a long hair for
a Schwarzschild~\cite{ach95} and Reissner-Nordstr\"om~\cite{cha98a,cha98b} black holes. The extremal Reissner-Nordstr\"om black hole
displays an analog of the Meissner effect, but the flux expulsion does not occur in all
cases~\cite{bon99}. The case of
a charged dilaton black hole Abelian Higgs vortex system was treated in~\cite{mod98a,mod99}, where among all it was shown that all
extremal dilaton black holes always expelled the vortex flux. Additionally,
the problem of a superconducting cosmic vortex and possible fermion condensations
around the Euclidean Reissner-Nordstr\"om~\cite{gre92} and dilaton~\cite{nak11} black holes were studied.
The physics of a black string with a superconducting cosmic string was treated in~\cite{nak12}. 

The case of a stationary axisymmetric black hole vortex system turns out to be far more complicated.
First attempts to attack this problem were presented in~\cite{ghe02}, but the correct
treatment of a Kerr black hole with an Abelian Higgs vortex was presented in~\cite{gre13}. The charged stationary axisymmetric solution of the low-energy limit of the heterotic string theory,
a Kerr-Sen black hole with an Abelian Higgs vortex was studied in~\cite{nak13}.

Recently, a proposal to look again into the old astrophysical observations like supernova 1987A data and to try to reinterpret them
taking into account the existence of the {\it dark radiation} (the dark photon) was delivered~\cite{cha17}. 
Moreover, some efforts to find strong constraints on emission of {\it dark photons} from stars~\cite{pos13} and on the coupling of {\it dark matter} coming from 
light particle production in hot star cores and their effects on star cooling were made~\cite{har17}. 
Moreover, a recent observational confirmation of the 
existence of gravitational waves renewed interest in analysis of the gravitational properties of cosmic string and domain walls networks~\cite{aze17,krajewski16}. 
Encouraged by this, we undertook studies of gravitational interactions and properties of the 
stationary axisymmetric black hole Abelian Higgs {\it dark matter} vortex system.
In our model
the Abelian Higgs sector was filled by
an additional $U(1)$-gauge field being the {\it dark photon} and a complex scalar representing a {\it dark matter} candidate.
The presence of the last two fields allowed us to take into consideration couplings between dark and visible sectors. These couplings were described by the 
Higgs portal interaction (quartic coupling between scalar fields) and a kinetic mixing type of coupling among gauge fields. 
Moreover, we also considered a possibility of the existence of nonminimal couplings between the scalars and gravity.         
The considered model was analyzed in the context of the existence of a string like solution in a flat spacetime in~\cite{hyd14}
and consequences of an existence of {\it dark strings} for the Standard Model particle emission were elaborated in~\cite{Long_Hyde_Vachaspati_2014,Long_Vachaspati_2014}.
Our work provides an expansion of research presented in~\cite{hyd14} to the case of strong gravitational field. 
We also hope that it will allow in the future to analyze the influence of strong gravitational field on various emission channels of the particles from the cosmic string.

The article is organized as follows. 
In section~\ref{perSol} we use the perturbative technique to solve the equations of motion for the Abelian Higgs model with the {\it dark matter} sector order by order. In the limit of a {\it thin string} we obtain an analytic expression for the line element
describing a stationary axisymmetric Kerr black hole with an Abelian Higgs {\it dark matter} vortex. 
In section~\ref{EOM} we shall not be interested in the backreaction of the cosmic string on the Kerr geometry but
 we focus on obtaining and analyzing properties of the field configurations for a subset of models discussed in section~\ref{perSol}.
The investigated model was chosen in such a way that it may represent a {\it dark photon} interacting with an ordinary photon via the kinetic mixing and with the Higgs sector via a complex scalar field with a quartic coupling to the Higgs field. 
Section~\ref{conclusions} contains a summary and discussion of the obtained results.

\section{Abelian Higgs model with the dark matter sector}
\label{perSol}
In this section we shall elaborate the problem of an Abelian Higgs {\it dark matter} vortex in the spacetime of a Kerr black hole solution.
The resulting action will be a sum of the Einstein gravity
\be
S_{gr} = {1 \over 16 \pi G}\int d^4x \sqrt{-g} \mathcal{R},
\ee
where $\mathcal{R} \equiv R^{\mu}_{~\mu}$ is the Ricci scalar. The action for an Abelian Higgs {\it dark matter} model with two $U(1)$-gauge fields coupled to complex scalars and 
possessing a kinetic mixing term between the gauge fields in question~\cite{ari14}
\ben
\label{action1}
S_{matter}&=& \int d^4x \sqrt{-g}\bigg[
- D^{(A)^\mu} \Phi^\dagger D^{(A)}_{\mu} \Phi - \frac{1}{4}F_{\mu \nu}F^{\mu \nu} - \frac{\tla}{4}
\left( \Phi^\dagger \Phi - \eta^2 \right)^2 + \\ \nonumber
&&- D^{(B)}_{ \mu} \Psi^\dagger D^{(B)\mu }\Psi - \frac{1}{4}B_{\mu \nu}B^{\mu \nu} - \frac{\tla_d }{4}
\left( \Psi^\dagger \Psi - \eta_d^2 \right)^2  - \frac{\kappa}{2}F_{\mu \nu}B^{\mu \nu} \bigg],
\een
where we denoted complex scalar fields by $\Phi$ and $\Psi$, coupled to $A_\mu$ and $B_\mu$ gauge fields, respectively. 
The adequate  field strength tensor bounded with $A_\mu$ is of the form
$F_{\mu \nu} = 2\na_{[\mu} A_{\nu]}$ and is connected with the visible sector, while 
the other $U(1)$-gauge field $B_\mu$ with the strength tensor $B_{\mu \nu} = 2\na_{[\mu}B_{\nu]}$ pertains to the {\it dark matter} sector.
The covariant derivatives charged under the visible and {\it dark matter} sectors are given by
$D^{(A)}_\mu = \na_\mu + ieA_\mu$ and $D^{(B)}_\nu = \na_\nu + ie_dB_\nu$, where $e_d$ is a charge of the {\it dark photon}.
The constants $\tla$ and $\tla_d$ denote the adequate couplings, while $\eta$ and $\eta_d$ are bounded with the energy scales at which
the $U(1)$-gauge symmetries responsible for the visible and connected with the {\it dark matter} sectors, respectively, are broken. Predicted values of the $\kappa$-coupling constant, being
the kinetic mixing parameter between the two $U(1)$-gauge fields, for realistic string compactifications range between $10^{-2}$ and $10^{-16}$ \cite{abe04}-\cite{ban17}.

One can define the aforementioned complex fields in terms of the real ones 
$X$, $Z_\alpha$, $\tchi$ and $Y$, $P_\alpha$, $\ttheta$ by the following relations:
\ben
\Phi(x_\mu) &=& \eta X(x_\mu) e^{i \tchi(x_\mu)},\\
\Psi(x_\mu) &=& \eta_d Y(x_\mu) e^{i \ttheta(x_\mu)},\\
A_\nu(x_\mu) &=& \frac{1}{e} \big[ Z_\nu(x_\mu) - \na_\nu \tchi(x_\mu) \big],\\
B_\nu(x_\mu) &=& \frac{1 }{e_d} \big[ P_\nu (x_\mu) - \na_\nu \ttheta(x_\mu) \big].
\een
The above real fields are associated with physical degrees of freedom for the broken symmetric phases, i.e., $X$, $Y$ are Higgs bosons in the visible and dark sectors, respectively, while
$Z_\alpha$, $P_\alpha$ are massive vector bosons. On the other hand, $\tchi$, $\ttheta$ are gauge degrees of freedom and they do not constitute local observables. The equations of motion for the above fields are provided by
\ben
\na_\mu \na^\mu X - Z_\mu Z^\mu X - \frac{\tla\eta^2}{ 2} \left( X^2 - 1 \right) X =& 0,\\
\na_\mu \na^\mu Y - P_\mu P^\mu Y - \frac{\tla_d\eta_d^2}{ 2} \left( Y^2 - 1 \right) Y =& 0,\\
\na_\mu \left( F^{\mu \nu} + \kappa B^{\mu \nu} \right) - 2 e^2\eta^2 X^2 Z^\nu =& 0,\\
\na_\mu \left( B^{\mu \nu} + \kappa F^{\mu \nu} \right) - 2 e_d^2\eta_d^2Y^2 P^\nu =& 0.
\een
In what follows we assume that 
\be
\tla_d = \xi_1 \tla, \qquad \eta_d = \xi_2 \eta.
\ee
The existence of vortex solutions in the Abelian Higgs case was discussed in \cite{nil73}. For the case when the loop encircles the vortex, the line integral of
$d \tchi$ and $d \ttheta$ around a closed loop (along which $X=Y=1$) is non-vanishing, i.e.,
$\int d \tchi = 2 \pi N_1$ and $\int d \ttheta = 2 \pi N_2$, where $N_a$ are the winding numbers. Continuity demands that at $X=Y=0$, at some point of any surface spanning the loop,
this is the locus of the vortex. 
Thus, the physical content of the considered model in question is related to the fields $P_\mu$, $Z_\mu$, $X$, $Y$, as well as, the  boundary conditions imposed on them.
The basic idea of looking for the vortex solution on the black hole is to obtain an approximate solution, having in mind that the string is much thinner than the event horizon of the black hole
\cite{ach95}-\cite{mod99}.
Namely, at large radii we should obtain the Nielsen-Olesen-like solution, while outside the event horizon of the black hole the solution ought to describe the core of the string.
Just, $(X,~Z_\mu) \rightarrow (1,~0)$ and $(Y,~P_\mu) \rightarrow (1,~0)$ when $r \rightarrow \infty$, while
$(X,~Z_\mu) \rightarrow (0,~1)$ and $(Y,~P_\mu) \rightarrow (0,~1)$ when $r \rightarrow r_h$.
Of course, as it will be explained in what follows, one should have in mind that the above statement is correct for the coordinate range where the string exists. For the case under consideration,
it does only for the polar coordinate $\theta \sim 0,~\pi$.

The idea is to solve the Einstein field equations with the Abelian Higgs {\it dark matter} energy momentum tensor. The field equations for the considered system yield
\be
G_{\mu \nu} = 8\pi GT_{\mu \nu}(vortex).
\ee
The energy momentum tensors for the adequate matter fields imply 
\ben \nonumber
T_{\mu \nu}(vortex) &=& 2\eta^2 \na_\mu X \na_\nu X + 2\eta^2 X^2Z_{\mu} Z_{\nu} + 
2\eta_d^2 \na_\mu Y \na_\nu Y + 2\eta_d^2 Y^2 P_{\mu} P_{\nu} + \\
&&+ F_{\mu \ga} F_{\nu}{}{}^{\ga} + B_{\mu \ga} B_{\nu}{}{}^{\ga}  + 2\kappa  B_{\mu \ga }F_{\nu}{}{}^{\ga}
+ g_{\mu \nu} \cL(\Phi,\Psi,A_\mu,B_\mu),
\een
where $\cL(\Phi,\Psi,A_\mu,B_\mu)$ is the Lagrangian density
\ben \nonumber
\cL(\Phi,\Psi,A_\mu,B_\mu) &=& - \eta^2 \na_\mu X \na^\mu X - \eta^2 X^2 Z_{\mu} Z^{\mu} -
 \eta_d^2 \na_\mu Y \na^\mu Y - \eta_d^2 Y^2 P_{\mu} P^{\mu} - \frac{1}{4} F_{\mu \nu} F^{\mu \nu} + \\
&& - \frac{\tla \eta^4}{ 4} \left( X^2 - 1 \right)^2 -
\frac{1}{4} B_{\mu \nu} B^{\mu \nu} - \frac{\tla_d \eta_d^4}{ 4} \left( Y^2 - 1 \right)^2 - \frac{\kappa}{2} B_{\mu \nu} F^{\mu  \nu}.
\een

In order to obtain the line element for the Abelian Higgs {\it dark matter} vortex in the spacetime of the stationary axisymmetric Kerr black hole,
we shall follow the approach presented in~\cite{ach95}. Namely, we use the perturbative technique and solve the underlying equations order by order in $\ep = 8 \pi G \eta^2$. 
The assumption of $\ep$ being small is well justified, e.g., for the grand unified string theory, $\ep \leqslant 10^{-6}$.
 The quantities will be expanded in series in order of $\ep$. As a background solution we take the Weyl form of the Kerr metric and find the exact form of the zero order quantities. Namely, the Weyl form of the axisymmetric line element is written in the form as
\be
ds^2 = - e^{2 \la_0} dt^2 + \alpha_0^2 e^{-2 \la_0} \left( d \phi + \beta_0 dt \right)^2 + e^{2(\nu_0 - \la_0)} \left( dx^2 + dy^2 \right),
\ee
where the functions under consideration are of $x$, $y$ dependences.
The corresponding Kerr line element implies
\ben \label{kerr}
ds^2 &=& -\left(1 -\frac{2GMr}{\trho^2} \right) dt^2 + \trho^2 \left( \frac{dr^2}{\Delta} + d\theta^2 \right)
+ \frac{4GMra\sin^2 \theta}{\trho^2} dt d\phi + \\ \nonumber
&&+ \bigg[ r^2 + a^2 + \frac{2GMra^2\sin^2\theta}{\trho^2} \bigg] \sin^2 \theta d\phi^2,
\een
where we have denoted
\be
\trho^2 = r^2 + a^2\cos^2 \theta, \qquad \Delta = r (r - 2 G M) + a^2.
\ee
$M$ is the standard Schwarzschild mass while $a$ is the Kerr
rotation parameter related to the black hole angular momentum by $a=\frac{J}{M}$. Defining $x$ and $y$ coordinates by the relations
\be
x = \int {dr \over \sqrt{\Delta}}, \qquad y = \theta,
\ee
one gets the following correspondence between the Kerr metric and the Weyl line element:
\be
\alpha_0 = \sqrt{\Delta}\sin\theta, \qquad
e^{2 \la_0} = {\trho^2 \Delta \over \Sigma^2}, \qquad
\beta_0 = - \frac{2MGar}{\Sigma^2}, \qquad
e^{2 \nu_0} = {\trho^4 \Delta \over \Sigma^2},
\ee
where $\Sigma^2 = \left(r^2 + a^2\right)^2 - a^2\Delta\sin^2\theta$.

We rewrite the Einstein equations in a more convenient form
\be
R_{\alpha \beta} = \ep T_{\alpha \beta} (vortex)- {1 \over 2}\ep T(vortex)g_{\alpha \beta}.
\ee
In the above relations $T_{\alpha}{}{}^\beta (vortex)$ is connected with the rescaled energy momentum tensor (see, e.g.,~\cite{dow92,mod99})
of the considered Abelian Higgs  {\it dark matter} vortex. In the 
discussed case of a stationary axisymmetric spacetime, one should also take into account
$A_{\phi}$ and $A_t$ components of the Abelian Higgs gauge field and the same components of the {\it dark matter} sector gauge field, i.e., $B_{\phi}$ and $B_t$.

In order to find the backreacted vortex solution,
one implements an iterative procedure of solving equations of motion, expanding the equations
in terms of $\ep$. 
To the zeroth order, our starting background solution will be described by $X_0$, $Y_0$ and $Z_0$, $P_0$,
and will constitute a kind of the Nielsen-Olesen ordinary and {\it dark matter} vortex solutions~\cite{nil73,ari14} provided by 
\ben
X \simeq X_{0}(R), \qquad Z_{\phi} \simeq Z_{0}(R), \\
Y \simeq Y_{0}(R), \qquad P_{\phi} \simeq P_{0}(R),
\een
respectively, where we have denoted by $R = \rho\sin\theta$. Because of the rotation one expects the occurrence of the mixing terms between $t$ and $\phi$ degrees of freedom \cite{gre13}.
Having this fact in mind we introduce the $t$-component of the considered gauge fields in the forms
\be
Z_{t} \simeq \zeta Z_{\phi}, \qquad P_{t} \simeq \zeta_d P_{\phi}.
\ee
It can be easily found that $\zeta = \zeta_d =-2MGar / \rho^4$, where we put $\rho$ equal to
$\rho^2 = r^2 + a^2$
and $R = \rho\sin\theta = \rho e^{- \la_0}$. Near the core of the cosmic string 
one has that $\sin \theta \sim \cO(M^{-1})$, which results in
the relation $R^2_{,x} + R^2_{,y} \sim e^{2\nu_0 - 2\la_0}$.
It all leads us to the conclusion that to the zeroth order in $\ep$ the components describing the energy momentum tensor
of the Abelian Higgs {\it dark matter} vortex yield
\ben \nonumber \label{t1}
T_{(0)y}^{}{}^{y} &\simeq& ({X'}_{0})^{2} + \xi_1^2({Y'}_{0})^{2} 
- \frac{X_{0}^2Z_{0}^2}{  R^2}
- \frac{\xi_2^2Y_{0}^2P_{0}^2}{  R^2}
+ {\tbeta_f}\frac{{Z'}_{0}^2}{  R^2} - {1 \over 4}\left( X_0^2 - 1 \right)^2 +\\
&&+ \frac{\tbeta_d}{\xi_1}\frac{{P'}_{0}^2}{  R^2} 
- \frac{\xi_1 \xi_2^4}{4}\left( Y_0^2 - 1 \right)^2 + \frac{\kappa\ga_{fd}}{R^2} P_{(0)}'Z_{(0)}', \\ \nonumber
T_{(0)\phi}^{}{}^{\phi} &\simeq& - ({X'}_{0})^{2} - \xi_1^2({Y'}_{0})^{2} + \frac{X_{0}^2 Z_{0}^2}{  R^2}
+ \frac{\xi_2^2Y_{0}^2P_{0}^2}{  R^2} 
({X'}_{0})^{2} - \xi_1^2({Y'}_{0})^{2}
- {1 \over 4}\left( X_0^2 - 1 \right)^2 + \\
&&- \frac{\xi_1 \xi_2^4}{4}\left( Y_0^2 - 1 \right)^2 + \frac{\kappa\ga_{fd}}{R^2} Z_{(0)}'P_{(0)}', \\ \nonumber
T_{(0)t}{}{}^{t} &=& T_{(0) x}{}{}^{x}
\simeq 
- ({X'}_{0})^{2} - \xi_1^2({Y'}_{0})^{2} - \frac{X_{0}^2Z_{0}^2}{R^2}
- \frac{\xi_2^2Y_{0}^2P_{0}^2}{R^2} - {\tbeta_f}\frac{{Z'}_{0}^2}{R^2} + \\
&&- \frac{\tbeta_d}{\xi_1}\frac{{P'}_{0}^2}{R^2} 
- \frac{\kappa\ga_{fd}}{R^2} Z_{(0)}'P_{(0)}', \\ \nonumber \label{tn}
T_{(0)xy} &\simeq& 2{\sqrt{\Delta} \over \rho}rR \bigg[
({X'}_{0})^{2} - \xi_1^2({Y'}_{0})^{2}  +
{\tbeta_f}\frac{{Z'}_{0}^2}{  R^2} - {1 \over 4}\left( X_0^2 - 1 \right)^2 + \\
&&+ \frac{\tbeta_d}{\xi_1}\frac{{P'}_{0}^2}{  R^2} + \frac{\kappa\ga_{fd}}{R^2} Z_{(0)}'P_{(0)}'\bigg], \\ \nonumber 
T_{(0)t}{}{}^{\phi} &\simeq& 0,
\een
where we have denoted
\ben
\tbeta_f = \frac{\tla}{2e^2}, \qquad \tbeta_d = \frac{\tla_d}{2e_d^2}, \qquad \ga_{fd} = \frac{\tla}{ee_d}, 
\een
and for the brevity of notation, in equations (\ref{t1})-(\ref{tn}), we change the definition of $R$, including in it $\tla$ and $\eta$, i.e.,
\be 
R = \sqrt{\tla}\eta\rho.
\ee
In what follows we shall use this definition.
A prime denotes differentiation with respect to the $R$-coordinate.

In the next step~\cite{ach95} we expand the adequate quantities in series, i.e.,
$\alpha = \alpha_0 + \ep\alpha_1$, etc., and solve the equations of motion iteratively.
Because of the fact that the energy momentum tensor components are functions of the $R$-coordinate one can draw a conclusion that
the modification of the Kerr stationary axisymmetric solution will also depend on this coordinate.
Consequently, we suppose that the first order perturbed solutions of the equations of motion of the underlying theory
will yield
\be
\alpha_1 = \alpha_0 \tal_1(R), \qquad \la_1 = \la_1(R), \qquad \nu_1 = \nu_1(R), \qquad \beta_1 = \beta_1(R).
\ee
The exact form of the energy momentum tensor for the cosmic vortex near the string core and the fact that
$(T_{t}{}{}^t + T_{\phi}{}{}^\phi - T)(X,Y,F,B)$ is equal to zero, 
lead us to the leading order equation of the form
\be
\frac{d^2 }{ d R^2} \tal_1 + \frac{2}{  R}\frac{d}{  d R} \tal_1 = 2{X_{0}^2Z_{0}^2 \over R^2}
+ \frac{1}{ 2}\left( X_0^2 - 1 \right)^2 
+ 2\frac{\xi_2^2Y_{0}^2P_{0}^2}{ R^2}
+ \frac{\xi_1\xi_2^4}{ 2}\left( Y_0^2 - 1 \right)^2.
\ee
It can be easily checked that $\tal_1(R)$ is given by
\be
\tal_1(R) = \int_R^\infty \frac{dR}{R^2}\int_0^R R'^{2} \bigg[
2\frac{X_{0}^2Z_{0}^2}{  R^2}
+ \frac{1}{ 2}\left( X_0^2 - 1 \right)^2 
+ 2\frac{\xi_2^2Y_{0}^2P_{0}^2}{ R^2}
+ \frac{\xi_1\xi_2^4}{ 2}\left( Y_0^2 - 1 \right)^2 \bigg] dR'.
\ee
On the other hand, the fact that the background function $\beta_0 =\cO(r^{-3})$ and its derivatives are subdominant quantities reveals that $\ep^1$-order Einstein equations have the forms
\ben \label{rff} \nonumber
R\left( R\tal_1^{''} + 2\tal_1^{'} - R\la_1^{''} - \la_1^{'} \right) =
R^2 \bigg[
2\frac{X_{0}^2R_{0}^2 }{R^2} + 2\frac{\xi_2^2Y_{0}^2P_{0}^2 }{ R^2}
+ {\tbeta_f}\frac{(R_{0}^{'})^2}{  R^2} 
+ \frac{\tbeta_d}{\xi_1}\frac{(P_{0}^{'})^2}{R^2}+ \\
+ \frac{\alpha\ga_{fd}}{R^2} R_{(0)}'P_{(0)}' 
+ \frac{1}{4}\left( X_0^2 - 1 \right)^2  + \frac{\xi_1\xi_2^4}{4}\left( Y_0^2 - 1 \right)^2 \bigg], \qquad
\een
\be \label{ryy}
\tal_1^{''} + 2\frac{\tal_1^{'}}{R} + 2\nu_{1}^{''} - 2\la_1^{''} - 2\frac{\la_1^{'}}{R} =
2(X_{0}')^2 + 2\xi_2^2(Y_{0}')^2  +
\frac{1}{2}\left( X_0^2 - 1 \right)^2 + \frac{\xi_1\xi_2^4}{2} \left( Y_0^2 - 1 \right)^2,
\ee
\ben \label{rxy} \nonumber
-\tal_1^{''} - \frac{\tal_1^{'}}{R} - \nu_{1}^{''} + \la_1^{''} + \frac{\nu_1^{'}}{R}
+ \frac{\la_1^{'}}{R} =
2(X_{0}')^2 + 2\xi_2^2(Y_{0}')^2  +
{\tbeta_f}\frac{(R_{0}^{'})^2}{R^2} + \\
+ \frac{\tbeta_d}{\xi_1}\frac{(P_{0}^{'})^2}{R^2} 
+  \frac{1}{4}\left( X_0^2 - 1 \right)^2 
 + \frac{\xi_1\xi_2^4}{4}\left( Y_0^2 - 1 \right)^2,
\een
\ben \nonumber
- \frac{\sqrt{\Delta}}{ \rho}r
\left( R^2\tal_1^{''} + 2R\tal_1^{'} - 4\la_1^{'} \right) \simeq
{\sqrt{\Delta} \over \rho}rR\bigg[
2(X_{0}')^2 + 2\xi_2^2(Y_{0}')^2 + \\
+ 2{\tbeta_f}\frac{(R_{0}^{'})^2}{  R^2} 
+ 2\frac{\tbeta_d}{\xi_1}\frac{(P_{0}^{'})^2}{R^2} 
+ 2\frac{\alpha\ga_{fd}}{R^2} R_{(0)}'P_{(0)}' \bigg].
\een

As in~\cite{mod98,ach95,cha98a,cha98b,bon99,mod98a,mod99}, we
shall work in the so-called {\it thin string limit}. It means that one assumes that
the mass of the black hole in question is subject to the inequality
$M \gg 1$ and one neglects terms of the order $\cO(1/M^{n\geqslant 2})$. 
Let us consider the equation (\ref{rff}) and the relation for $\talpha_1$. It provides the following expression for $\la_1$:
\ben \nonumber
\la_1 \simeq \int_R^\infty {dR \over R}\int_0^R dR'{R'}\bigg[
\frac{1}{ 4}\left( X_0^2 - 1 \right)^2 + \frac{\xi_1\xi_2^4}{ 4}\left( Y_0^2 - 1 \right)^2 
- {\tbeta_f}\frac{(R_{0}^{'})^2}{ {R'}^2} + \\
- \frac{\tbeta_d}{\xi_1}\frac{(P_{0}^{'})^2}{  R^2}  - \frac{\alpha\ga_{fd}}{R^2} R_{(0)}'P_{(0)}' \bigg].
\een
From the relations \eqref{ryy} and \eqref{rxy} one has that $\nu_1 = 2\la_1$. 

As far as the $\beta_0$ correction is concerned, the first order cannot be determined by an asymptotical analysis.
It is caused by the fact that the quantity $\beta_0$ is a subdominant function in the
problem in question. Taking the divergent part of the Ricci curvature tensor $R_{xy}$, we arrive at the relation
\be
e^{-4\la_0}\alpha_0^2\beta_{0,x}\beta_{1,y} \simeq
{2MaR^2 \over \rho^2\sqrt{\Delta}} \bigg(
-1 + \frac{4r^2 }{ \rho^2} + \cO \bigg( {1 \over \rho^4} \bigg)\bigg)\beta_{1,y}.
\label{bet}
\ee
It is customary to elaborate the right-hand side of \eqref{bet} 
examining the adequate components of the energy momentum tensor, both for the Abelian Higgs {\it dark matter} vortex.
One deduces that we cannot find $\delta \beta_0 = \beta_1$ which has the
mandatory functional dependence on the coordinates.
In the light of this statement, the required form of the $\ep^1$-order $\beta_0$ correction required for a pure
$\phi$ deficit angle leads to the divergence of the $xy$-component of the Ricci curvature tensor
at the event horizon of the Kerr black hole. In the view of these arguments, we deduce that 
$\delta \beta_0 = 0$.

Rescaling the coordinates
$\hht = e^{\ep\nu_1/2}t$, $\hr = e^{\ep\nu_1/2}r$, $\ha = e^{\ep\nu_1/2}a$, $\hM = e^{\ep\nu_1/2}M$, one arrives at the line element describing the {\it thin dark string} in the Kerr black hole spacetime.
One can notice that the presence of the vortex modified the black hole parameters as angular momentum and mass, measured at infinity by the adequate factor. The same situation was revealed
in the case of votex static black hole systems \cite{ach95}-\cite{mod99}, as well as, for stationary axisymmetric string systems \cite{gre13,nak13}.\\
The metric of the Kerr black hole threaded by a cosmic {\it dark} string is provided by
\ben \nonumber
ds^2 &=& - \left( {\hDelta - \ha^2 \sin^2 \theta \over \hrho^2} \right) d\hht^2 
- {8G^2\hM^2\ha^2\hr^2 \over \hSigma^4}\ep\hmu \left(\hr^2+ \ha^2 \right) \sin^2\theta d \hht^2 
+ {\hSigma^2 \over \hDelta} d\hr^2 + \hrho^2 d\htheta^2 +\\
&&+ {\hSigma^2\sin^2 \theta \over \hrho^2} \left( 1 - 2\ep\hmu \right) d\phi^2 
- \frac{4G\hM\ha\hr}{\hrho^2}\sin^2\theta \left( 1 - 2\ep\hmu \right) d\hht d\phi,
\een
where $\hmu$ is the mass per unit length of the cosmic string~\cite{dow92}, while 
the other quantities are defined as follows:
\ben
\hDelta &=& \hr (\hr - 2G\hM) + \ha^2,\\
\hrho^2 &=& \hr^2 +\ha^2\cos^2\theta,\\
\hSigma^2 &=& \Big( \hr^2 +\ha^2 \Big)^2 - \ha^2\hDelta^2\sin^2 \theta.
\een
As far as $\hmu$ is concerned, it has the following form:
\be
\hmu = \hmu_{ord} + \hmu_{dm},
\ee
where we have denoted by $\hmu_{ord}$ the mass of the visible matter sector per unit length of the cosmic string
\be
\hmu_{ord} \simeq \int dRR\bigg[
\frac{X_{0}^2 Z_{0}^2 }{ R^2} + \frac{1}{ 4} \left( X_0^2 - 1 \right)^2 + \tbeta_f \frac{(Z_{0}')^2}{R^2} \bigg],
\ee
while for $\hmu_{dm}$ we set
\be
\hmu_{dm} \simeq \int dRR\bigg[
\frac{\xi_2^2Y_{0}^2P_{0}^2 }{ R^2} + \frac{\xi_1\xi_2^4}{ 4}\left( Y_0^2 - 1 \right)^2 + \frac{\tbeta_d}{\xi_1}\frac{(P_{0}^{'})^2}{R^2} 
+ \frac{\kappa\ga_{fd}}{R^2} Z_{(0)}' P_{(0)}' \bigg].
\ee
As was noticed in~\cite{gre13} and confirmed for the Kerr-Sen vortex system~\cite{nak13}, the Abelian Higgs {\it dark matter} vortex also causes not only 
an angular deficit angle but the deficit angle affects the time coordinate. The new effect is the influence on the ergosphere because of the fact that
the deficit angle appears in $g_{tt}$ component of the metric tensor, which constitutes the condition for the ergosphere radius, i.e., $g_{tt}(r_{erg})=0$. The position of the ergosphere is shifted.
In the considered case we obtain a larger destruction of the ergosphere comparing to the Kerr-Abelian Higgs vortex system, due to the fact that $\hmu = \hmu_{ord} + \hmu_{dm}$.
On the other hand, the limiting angular velocity at the Kerr black hole threaded by the {\it dark matter} vortex yields
\be
\Omega_h = \omega^2_{Kerr}\bigg[ 1 + \ep\left(\hmu_{ord} + \hmu_{dm}\right) \left(2 - 8G\hM\hr \right) \bigg],
\ee
where $\omega^2_{Kerr} = {\ha^2 / \left(\hr^2 + \ha^2\right)}$. It is composed of the velocity associated with the existence of the Abelian-Higgs vortex plus the part influenced by the {\it dark matter} sector
and the part responsible for the interaction of the ordinary matter field with {\it dark matter}. Just
\be
\Omega_h (Kerr-dm~vortex) > \Omega_h (Kerr-vortex).
\ee
The same situation holds if one considers the area of the black hole - vortex event horizon. Namely,
up to the $\ep^1$-order one gets
\be
A_{Kerr-dm~vortex} \sim A_{Kerr} \sqrt{1 - 2\ep (\hmu_{ord} + \hmu_{dm})}.
\ee
The horizon area of the Kerr - {\it dark matter} vortex system is smaller than the area of the event horizon for the Kerr-Abelian Higgs vortex system and the Kerr black hole
\be
A_{Kerr-dm~vortex} < A_{Kerr-vortex} <A_{Kerr}.
\ee

\section{Cosmic string in the Higgs -- scalar mediator -- heavy dark photon sector }
\label{EOM}
Next, we will focus on the subspace of the models discussed in the previous section which are 
closer to what is investigated in the context of particle physics (see, e.g.,~\cite{hyd14,Long_Hyde_Vachaspati_2014,Long_Vachaspati_2014}).
To this end, we retain the Abelian Higgs model for the {\it dark matter} sector with the kinetic coupling between the {\it dark photon} and the ordinary electromagnetic field.
We restrict our considerations to the case when
the second complex scalar field in the action~(\ref{action1}) is real. This implies that it may represent the Standard Model 
Higgs field (precisely, the real component of the Higgs field in a unitary gauge). Further,  we examine a quartic coupling between the Higgs and the complex scalar.
In order to make the results more transparent we begin with writing the action and discussing the field content of the model in question.

The cosmic string is composed of the {\it dark} sector fields within the Abelian Higgs 
model~(\ref{Scs}), while ordinary matter is given by the electromagnetic field and a real scalar field representing the real 
component of the Higgs field in the unitary gauge~(\ref{SHiggsEM}). Additionally, we consider both scalar fields possessing two types of coupling between the {\it dark matter} and ordinary matter sectors. 
These coupling channels are known as the kinetic mixing (coupling between the electromagnetic 
field and the {\it dark photon}) and the Higgs portal (quartic coupling between the Higgs field and the complex scalar). The appropriate part of the action is given by~(\ref{Scoupling}).
\begin{align}
\label{Scs}
S_{cs} &= \int \sqrt{-g} d^4 x \left[
- \frac{1}{4} B_{\mu \nu} B^{\mu \nu} -  D^{B}_{\mu} \Psi^{*} D^{B \mu} \Psi - \frac{\tilde{\lambda}_d}{4} \left( \Psi^{*} \Psi - \eta_d^2 \right)^2 \right], \\
\label{SHiggsEM}
S_{Higgs + EM} &= \int \sqrt{-g} d^4 x \left( - \frac{1}{4} F_{\mu \nu} F^{\mu\nu}
- \frac{1}{2} \nabla_{\mu} h \nabla^{\mu} h + \frac{1}{2} m_h^2 h^2 - \frac{\lambda_{h}}{4} h^4 \right), \\
\label{Scoupling}
S_{cs-int} &= \int \sqrt{-g} d^4 x \left(
-\frac{\kappa}{2} B_{\mu \nu} F^{\mu \nu}  - \lambda_{hY} \Psi^{*} \Psi h^2 \right),
\end{align}
where the dark matter fields $B_{\mu}$ and $\Psi$ and the electromagnetic field $A_{\mu}$ are defined as in section \ref{perSol}.
Meanwhile, we restrict the second scalar field to be real and reparametrize its potential. To conform our notation to the standard one used in 
particle physics we label this real field by $h$. $m_h$ represents the mass parameter for the Higgs field and $\lambda_{hY}$ is the 
quartic coupling between $\Psi$ and $h$. 
As a side note, we want to stress that since the only gauge boson of the Standard Model considered by us is the photon, 
we may represent the Higgs field as a single real scalar field $h$.

As far as the gravity part is concerned, we restrict our attention to the standard Einstein-Hilbert action. 
The background spacetime metric will be given by an asymptotically flat, stationary axisymmetric solution of vacuum Einstein equation (Kerr black hole)~(\ref{kerr}) or a static, spherically symmetric one
(Schwarzschild black hole). The latter can be obtained from~(\ref{kerr}) by the substitution $a=0$. $M$ is the black hole mass and the spin parameter of the black hole is given by
$a=\frac{J}{M}$, where $J$ is the black hole angular momentum.

In order to derive the equation of motion for the matter content we made the following substitutions for the {\it dark sector} fields:
$B_{\mu} = \frac{1}{e_d} \left( P_{\mu} - \nabla_{\mu} \tilde{\theta} \right)$ and $\Psi = \eta_d Y e^{i \tilde{\theta}}$. This replacement  amounts to the choice of the 
representation of the complex scalar field in terms of the two real fields $Y$ and $\tilde{\theta}$. Moreover, we will be working in the Lorentz
gauge $\nabla^{\mu}B_{\mu} = 0$, which implies $\nabla^{\mu} P_{\mu} = 0$ provided $\square \tilde{\theta} = 0$, where $\square$ is the covariant d'Alembert operator.
If the scalar field $\Psi$ forms a cosmic string the last condition is trivially satisfied since $\tilde{\theta} = N \phi$, where $\phi$ is the coordinate 
representing the azimuthal angle in the spherical coordinate system.
Furthermore, to simplify the equations of motion we diagonalized the gauge field kinetic terms by the transformation $C_{\mu} = A_{\mu}  + \kappa B_{\mu}$,
which is always possible for the Abelian gauge fields. The action for the matter sector after the discussed transformations implies
\begin{align}
S_{matter} =& S_{cs} + S_{Higgs+EM} + S_{cs-int} = \nonumber \\
=& \int \sqrt{-g} d^4x \bigg \{
- \frac{1}{4} C_{\mu \nu} C^{\mu \nu} - \frac{1}{4} \frac{ \tilde{\kappa}}{ e_d^2} P_{\mu \nu} P^{\mu \nu} + \nonumber \\ 
&+ \eta_d^2 \left[ 
- \nabla_{\mu} Y \nabla^{\mu} Y - Y^2 P_{\mu} P^{\mu} - \frac{\tilde{\lambda}_d \eta_d^2}{4} \left( Y^2 -1 \right)^2
- \lambda_{hY} h^2 Y^2  \right] + \nonumber \\
&- \frac{1}{2} \nabla_{\mu} h \nabla^{\mu} h  + \frac{1}{2} m_h^2 h^2 - \frac{\lambda_h}{4}h^4  \bigg \},
\end{align}
where we have set $C_{\mu \nu} = 2 \nabla_{[ \mu} C_{\nu ]} $ and $P_{\mu \nu} = 2 \nabla_{[ \mu} P_{\nu ]}$.
For further analysis it will be advantageous to regroup various terms in the above action. One thus obtains
\begin{align}
S_{matter} =& \int \sqrt{-g} d^4x \bigg \{
- \frac{1}{4} C_{\mu \nu} C^{\mu \nu} 
- \frac{1}{2} \nabla_{\mu} h \nabla^{\mu} h  + \frac{1}{2} m_h^2 h^2 - \frac{\lambda_h}{4}h^4  
 +  \nonumber \\
&+ \eta^2 \left[ 
- \frac{1}{4} \frac{2 \tilde{\kappa}}{ m_V^2} P_{\mu \nu} P^{\mu \nu} 
- \nabla_{\mu} Y \nabla^{\mu} Y - Y^2 P_{\mu} P^{\mu} - \frac{ m_Y^2}{4} \left( Y^2 -1 \right)^2 +\right. \nonumber \\
&- \lambda_{hY} h^2 Y^2  \bigg] \bigg \},
\end{align}
where the notation $m_{Y}^2 = \tilde{\lambda}_{d} \eta_d^2$, $m_{V}^2 = 2 e_d^2 \eta_d^2$ and $\tilde{\kappa} = 1 - \kappa^2$ was introduced.
In what follows, we will be interested only in the electromagnetic field induced by the {\it dark sector}, which implies $C_{\mu} = 0$ (see equations of motion (\ref{eomC})).
In the absence of the cosmic string we have $P^{\mu} = 0$ and one of the possible solutions to the equations of motion for the scalars is of the form
\begin{align}
\label{NoStringMin}
Y^2_{no string} &= \frac{ m_Y^2 \lambda_h - 2 m_h^2 \lambda_{hY}}{m_Y^2 \lambda_h - 4 \eta_d^2 \lambda_{hY}^2}, \nonumber \\
h^2_{no string} &= \frac{m_h^2 m_Y^2 - 2 \eta_d^2 m_Y^2 \lambda_{hY}}{m_Y^2 \lambda_h - 4 \eta_d^2 \lambda_{hY}^2}.   
\end{align} 
This solution implies that the scalars lie in the minimum of their potential.

Since we will be solving the equations of motion numerically, it is convenient to introduce a set of proper dimensionless quantities.
In our case this means the following redefinitions of the radial coordinate and black hole parameters:
\begin{align}
\label{subBH}
\frac{a}{GM} \rightarrow a, \quad \frac{r}{GM} \rightarrow r, \quad \frac{GM}{GM} \rightarrow 1,
\end{align}   
as well as fields and various parameters in the matter sector
\begin{align}
\label{subFields}
&GM m_Y \rightarrow m, \quad
GM m_h = \frac{GM m_h}{GM m_Y} GM m_Y \rightarrow c m, \quad  G^2 M^2 m_{V}^2 \rightarrow \frac{m^2}{\tilde{\beta}_d}, \nonumber \\
&GM h \rightarrow h, \quad (GM)^2 P^{\phi} \rightarrow P^{\phi}, \quad GM P^{t} \rightarrow P^{t},
\end{align} 
where $c$ is a number and $\tilde{\beta}_{d} = \frac{m_{Y}^2}{m_{V}^2}$. As can be seen from the last substitution in (\ref{subBH})
and the second one from (\ref{subFields}), the change in the value of the parameter $m$ may be prescribed to the change of the 
additional scalar mass or black hole mass. 
From now on, in order to get rid of this ambiguity, 
we fix the {\it dark scalar} mass to be constant. Its exact value will be discussed in the next subsection.
This implies that various values of $m$ will represent various black hole masses. The choice of the particular numeric value of $m_{Y}$ is arbitrary. 
We partially lift this arbitrariness by a demand that $Y$ could be a viable dark matter candidate~\cite{guo10,PhysRevD.88.055025,Feng2015}.

Finally, below we present the equations of motion for the matter fields expressed in terms of the dimensionless fields and parameters. They yield
\begin{align}
\label{eomX}
&\square Y - P^2 Y - \frac{1}{2}m^2 Y (Y^2 -1) - \lambda_{hY} h^2 Y = 0, \\
\label{eomh}
&\square h + c^2 m^2 h - \lambda_h h^3 - 2 \frac{\lambda_{hY}}{\tilde{\lambda}_d} m^2 Y^2 h = 0, \\
\label{eomP}
&\square P^{\alpha} - \frac{m^2}{\tilde{\kappa} \tilde{\beta}_d} Y^2 P^{\alpha} - R^{\alpha}_{~\beta} P^{\beta} = 0, \\
\label{eomC}
&\nabla_{\alpha}C^{\alpha \beta} = 0,
\end{align}
where $P^2$ and the $(r,\theta)$ part of the d'Alembertian operator expressed in the dimensionless quantities are given as follows:
\begin{align}
&\square = \frac{\Delta}{\trho^2} \partial^2_{r} + 
 2 \frac{r - 1}{\trho^2}  \partial_{r} 
+ \frac{1}{\trho^2} \partial^{2}_{\theta} + \frac{1}{\trho^2} \cot \theta  \partial_{\theta}, \\
&P^2 = \frac{\Sigma^2 \sin^2 \theta}{ \trho^2} (P^{\phi})^2 - 
\frac{ \Delta - a^2 \sin^2 \theta}{\trho^2} (P^{t})^2 -
\frac{ 4 a r \sin^2 \theta}{\trho^2} P^{t} P^{\phi},
\end{align}
under the condition that 
both $P^{\phi}$ and $P^{t}$ are nonzero. In the derivation of the equation of motion for $P_\mu$ field we use the Ricci identity and impose the Lorentz gauge
$\nabla_{\mu} P^{\mu} = 0$.

 Let us point out that far away from the cosmic string and the black hole we demand that scalar fields should approach their constant values given by the equation (\ref{NoStringMin}),
therefore it is convenient to rewrite them in terms of the dimensionless parameters  as follows: 
\begin{align}
\label{fieldsNoString}
h^2_{nostring} = \frac{ \tilde{\lambda}_d c^2 m^2 - 2 \lambda_{hY} m^2}{\lambda_h \tilde{\lambda}_d - 4 \lambda_{hY}^2}, \nonumber \\
Y^2_{nostring} = \frac{\lambda_h \tilde{\lambda}_d - 2 \lambda_{hY} \tilde{\lambda}_d c^2}{\lambda_h \tilde{\lambda}_d - 4 \lambda_{hY}^2}.
\end{align}

\subsection*{Numerical solution of the equations of motion }
In this subsection we solve numerically the system of equations (\ref{eomX})-(\ref{eomP}). 
As far as the equation (\ref{eomC}) is concerned, we will be interested in the solution $C^{\mu} = 0$.  The definition of $C^{\mu}$ ($C^{\mu} = A^{\mu} + \kappa B^{\mu}$)
implies that the electromagnetic field is induced by the presence of the {\it dark photon} field $B^{\mu}$. 
Moreover, the obtained form of $B^{\mu}$ fixes the functional form of $A^{\mu}$. 
The relaxation technique~\cite{Press:2007:NRE:1403886} will be implemented to find a numerical solution of the equations of motion describing the system in question.

In order to solve the system of elliptic equations one demands a specification of boundary conditions. In the case of the cosmic string in the black hole background, the
boundaries of the numerical domain are located at the string core, namely $\theta = 0$ and $\theta = \pi$, where $\theta$ is the polar angle in the spherical coordinate
system, and at some arbitrary selected large radius $r_{max}$. The $r_{max}$ is chosen to be much bigger than the other length scale present in the system, namely the black hole 
event horizon. In our calculations we assume $r_{max} = 25 r_{+}$, where $r_{+}$ is the event horizon of the Kerr or Schwarzschild black hole.
The last boundary of the numerical domain is located at the event horizon of the considered black hole $r_{+}$. In contrast to the other boundaries, the standard treatment 
of the cosmic string black hole configuration is not to impose the prescribed boundary value for the fields at $r_{+}$, but to solve the reduced form of the system in question.
By the reduced system we mean equations~(\ref{eomX})-(\ref{eomP}) in the limit $r=r_{+}$, which are solved for $Y(\theta)$, $P^{\mu}(\theta)$, $h(\theta)$.
After explaining the general form of the boundary conditions needed for the field configuration of the studied system, let us turn to the specific values used in our numerical calculations.

At the string core, $\theta = 0, \pi$, we put the usual boundary condition for $P^{\phi}$, namely $P^{\phi} = 1$. Since the $R^{\alpha}_{~\beta}$ is zero for the Kerr black hole
we may put $P^{t} = 0$. From the equation~(\ref{eomP}) we may infer that $P^t$ will not be sourced by other components of the vector field $P^{\mu}$. 
As far as the scalar sector is concerned, on the string we have $Y = 0$ which implies $h^2 = \frac{c^2 m^2}{\lambda_h}$.
The boundary conditions for large radii (but for $\theta \neq 0, \pi$ ) are given by $P_{r_{max}}^{\phi} = 0$, 
$Y_{r_{max}}^2 = Y^2_{nostring}$, $h_{r_{max}}^2 = h^2_{nostring}$, where $Y^2_{nostring}$ and $h^2_{nostring}$ are defined in~(\ref{fieldsNoString}).
It is straightforward to see that if we put $\lambda_{hY} =0$ we 
decouple $Y$ and $h$ fields and obtain $Y_{r_{max}} = 1$, which is the usual value of the $Y$ field far away from the string core.

After the discussion of the general form of the employed boundary conditions, let us turn to the discussion of the choice of the free parameters. 
We start by defining physical masses of the particles. In the flat spacetime at the tree-level (at the classical level) physical masses of the
particles could be defined as the eigenvalues of the mass matrix. Meanwhile, the mass matrix for the system of the scalar fields is defined as 
the matrix of the second derivatives of the potential calculated at its minimum. For the discussed case the potential part of the
Lagrangian for the scalar sector is provided by (for now, we return to the dimensionfull quantities):
\begin{align}
\label{Vfields}
V = \eta_d^2 Y^2 P_{\mu}P^{\mu} + \frac{\tilde{\lambda}_d}{4} \left(\eta_d^2 Y^2 - \eta_d^2 \right )^2 + \lambda_{hY} h^2 \eta_d^2 Y^2 
 - \frac{m_{h}^2}{2} h^2 + \frac{\lambda_h}{4}h^4.
\end{align}
At the minimum of the potential the following relations are fulfilled:
\begin{align}
\label{VminX}
&2 P_{\mu}P^{\mu}  + \tilde{\lambda}_d \left ( \eta_d^2 Y^2 - \eta_d^2 \right ) + 2 \lambda_{hY} h^2 = 0, \\
\label{Vminh}
&2 \lambda_{hY} \eta_d^2 Y^2 - m_h^2 + \lambda_h h^2 = 0. 
\end{align}
As we wrote earlier, the mass matrix is defined as $M^2 \equiv \begin{bmatrix} \partial^{min}_{\eta_d Y \eta_d Y} V & \partial^{min}_{\eta_d Y h} V \\
\partial^{min}_{h \eta_d Y} V & \partial^{min}_{ h h}V
\end{bmatrix} $,
the superscript $min$ means that the given quantity is calculated at the minimum of $V$ which implies that relations (\ref{VminX})-(\ref{Vminh}) hold
and we differentiated $V$ with respect to $\eta_d Y$ instead of $Y$ to obtain the correct dimension of the appropriate terms.
For the discussed case the mass matrix is given by
\begin{align}
\label{M2f}
M^2 = \begin{bmatrix}
2 \tilde{\lambda}_d v_Y^2 & 4 \lambda_{hY} v_h v_Y \\
4 \lambda_{hY} v_h v_Y & 2 \lambda_h v_H^2
\end{bmatrix}.
\end{align}
In the above expression we replaced the fields by their vacuum expectation values (vevs) $h \rightarrow v_h$ and $\eta_d Y \rightarrow v_Y$.
Since in the considered case $h$ represents the real component of the Higgs field, we fix $v_h = 246.2 \textrm{GeV}$. In the particle physics literature the 
vev of the second scalar is usually parametrized by the quantity $\tan{\tilde{\beta}} = \frac{v_h}{v_Y}$, for the purpose of numerical computations we fix it to be $\tan{\tilde{\beta}} = 0.001$. 
At this point let us focus on the question of neglecting the backreaction of the cosmic string on the geometry. For the chosen 
values of parameters $\tan{\tilde{\beta}}$ and $v_h$ the parameter $\epsilon$ governing the backreaction
is of the order of $10^{-26}$. This justifies neglecting the effect of the backreaction. 
The squares of physical masses of scalars are given by the eigenvalues of the mass matrix (\ref{M2f}), let us call them $m_{\pm}^2$. 
In the considered scenario we assumed that the second scalar is heavier than the Higgs. Taking this into account, we choose $m_{-} = 125.5 \textrm{GeV}$ and $m_{+} = 900 \textrm{GeV}$. 
At this point it is worth to count the number of parameters characterizing the scalar sector.
Far away from the cosmic string we have by definition of the cosmic string $P^{\mu} = 0$.
This implies that the scalar sector is effectively described by five parameters $\lambda_h$, $\tilde{\lambda}_d$, $\lambda_{hY}$, $m_h$, $m_Y$. So far we have specified four parameters
$v_h$, $\tan{\tilde{\beta}}$, $m_{-}$, $m_{+}$. The last remaining parameter is called the mixing angle $\alpha_M$ and is given by the relation as follows:
\begin{align}
\begin{bmatrix}
H_{-} \\ H_{+}
\end{bmatrix} = 
\begin{bmatrix}
\cos{\alpha_M} & - \sin{\alpha_M} \\
\sin{\alpha_M} & \cos{\alpha_M}
\end{bmatrix}
\begin{bmatrix}
h \\ \eta_d Y 
\end{bmatrix}, 
\end{align}
where $H_{\pm}$ are eigenstates of the $M^2$ matrix with eigenvalues $m^2_{\pm}$, respectively. Physically, a nonzero value of $\sin{\alpha_M}$
means that the physical Higgs field $H_{-}$ with mass $m_{-}$ is some mixture of the $h$ and $Y$ fields. In the totally decoupled case ($\sin{\alpha_M} = 0$)
we have $H_{-} = h$ and $H_{+} = \eta_d Y$. In our calculations we fix $\alpha_M$ to be $\sin{\alpha_M} = 0.023$. 
The particular values of parameters 
$v_h = 246.2 \textrm{GeV}$, $\tan{\tilde{\beta}} = 0.001$, $m_{-} = 125.5 \textrm{GeV}$, $m_{+} = 900\textrm{GeV}$, $\sin{\alpha_M} = 0.023$ 
imply that the $Y$ field could be a viable dark matter candidate~\cite{Robens2015,Feng2015}. 
Translations of this set of parameters to the one present in the potential (\ref{Vfields}) give us the following relations (see, e.g.,~\cite{Robens2015}):
\begin{align}
\lambda_h &= \frac{ m_{-}^2}{2 v_h^2} \cos^2{\alpha_M} + \frac{m_{+}^2}{2 v_h^2} \sin^2{\alpha_M}, \\
\lambda_X &= \frac{ m_{-}^2}{2 v_Y^2} \sin^2{\alpha_M} + \frac{m_{+}^2}{2 v_Y^2} \cos^2{\alpha_M}, \\
\lambda_{hX} &= \frac{m_{+}^2 - m_{-}^2}{2 v_h v_Y} \sin{2 \alpha_M}.
\end{align}  
As far as the $m_h$ and $m_Y$ are concerned, they also could be expressed by the same physical parameters as $\lambda$s 
yet their numerical values do not appear in our equations, what is evident from the rescaling (\ref{subFields}).
On this account, we present here only the numerical value of the ratio of the masses $c = \frac{m_h}{m_Y} = 9.5$ and 
point out again that $m_Y$ is encompassed in the dimensionless parameter $m$.
To summarize, during our calculations we fix the following dimensionless parameters present in the equations 
(\ref{eomX})-(\ref{eomC}):
\begin{align}
\tilde{\beta}_d = 1, c = 9.5, \lambda_h = 0.133, \tilde{\lambda}_d = 6.679 \cdot 10^{-6}, 
\lambda_{hY} = 3.013 \cdot 10^{-4}, \kappa = 10^{-5},
\end{align}
where the last parameter enters the equation (\ref{eomP}) through the relation $\tilde{\kappa} = 1 - \kappa^2$ and represents the strength
of the kinetic mixing among the gauge fields. 
As far as the masses of the black holes are concerned, for the specified and fixed $m_{+}$ (which gives fixed $m_{Y}$) we have
that $m = 1$ implies a black hole of a mass $M_{BH} \sim 2 \cdot 10^{11} g \sim 10^{-22} M_{\odot}$.

As the last step in preparation to discuss our results, let us restrict our attention to the ordinary electromagnetic field $A_{\mu}$.
We are interested in the trivial solution to (\ref{eomC}), namely $C^{\mu}=0$, but this
does not imply a trivial $A_{\mu}$ field. Using the relations $C^{\mu} = A^{\mu }+ \kappa B^{\mu}$, 
$B^{\mu} = \frac{1}{e_d} \left( P^{\mu} - \nabla^{\mu} \tilde{\theta} \right)$ and $\tilde{\theta} = N \phi$ we obtained
\begin{align}
\label{EM_potencjal}
A_t &= - \frac{\kappa}{e_d} \left( g_{tt} P^t + g_{t \phi} P^{\phi} \right), \\
\label{EM_potencjal2}
A_{\phi} &= - \frac{\kappa}{e_d} \left( g_{\phi \phi} P^{\phi} + g_{t \phi} P^{t} - 2 N \right),
\end{align} 
where the charge $e_d$ can be expressed by the dimensionless parameters as $e_d^2 = \frac{\tilde{\lambda}_d}{2 \tilde{\beta}_{d}}$.

\begin{figure}[H]
\centering
\subfloat[ ]{
  \includegraphics[width=.46\textwidth]{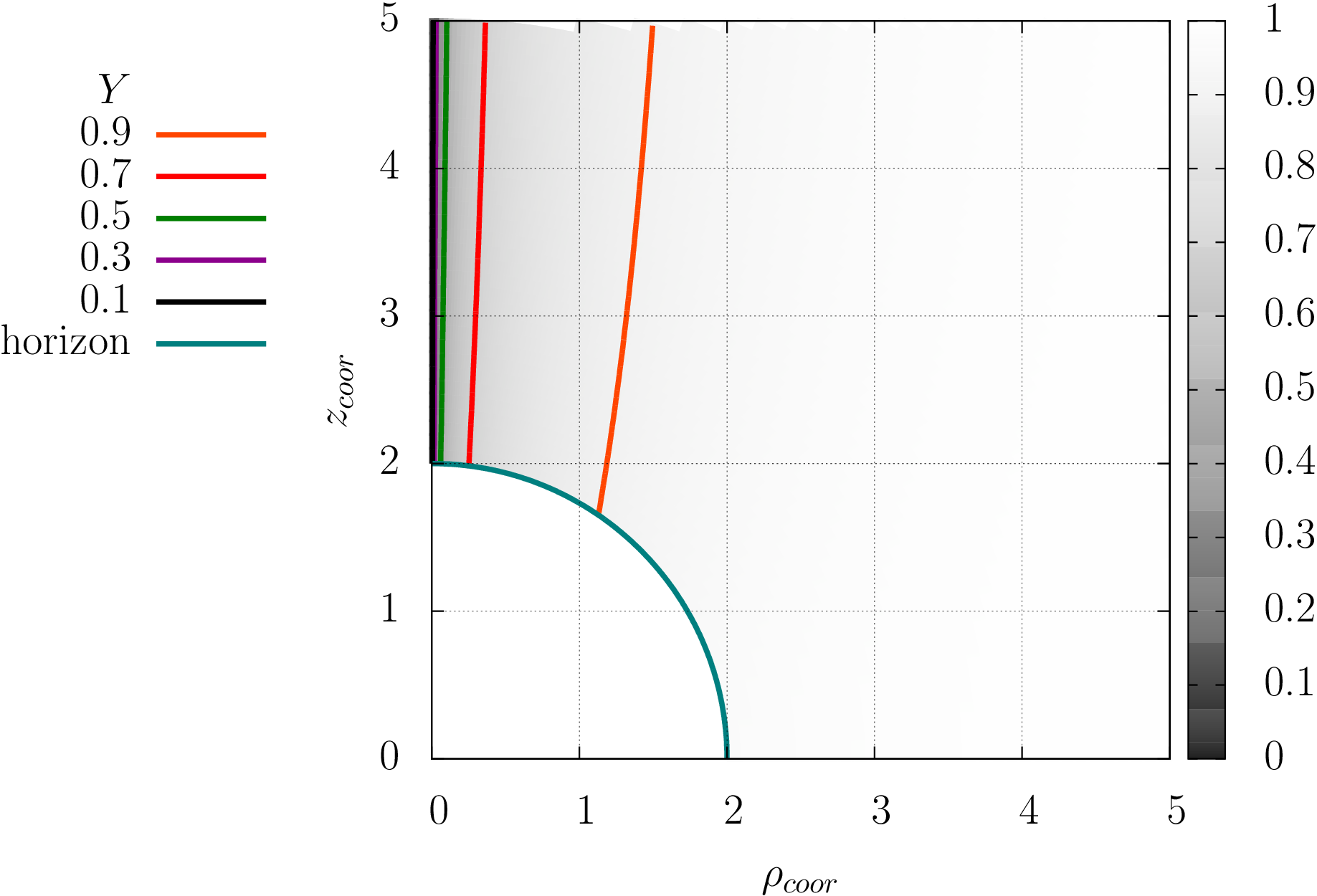}
\label{fig1a}
  }
\quad
\subfloat[]{
  \includegraphics[width=.46\textwidth]{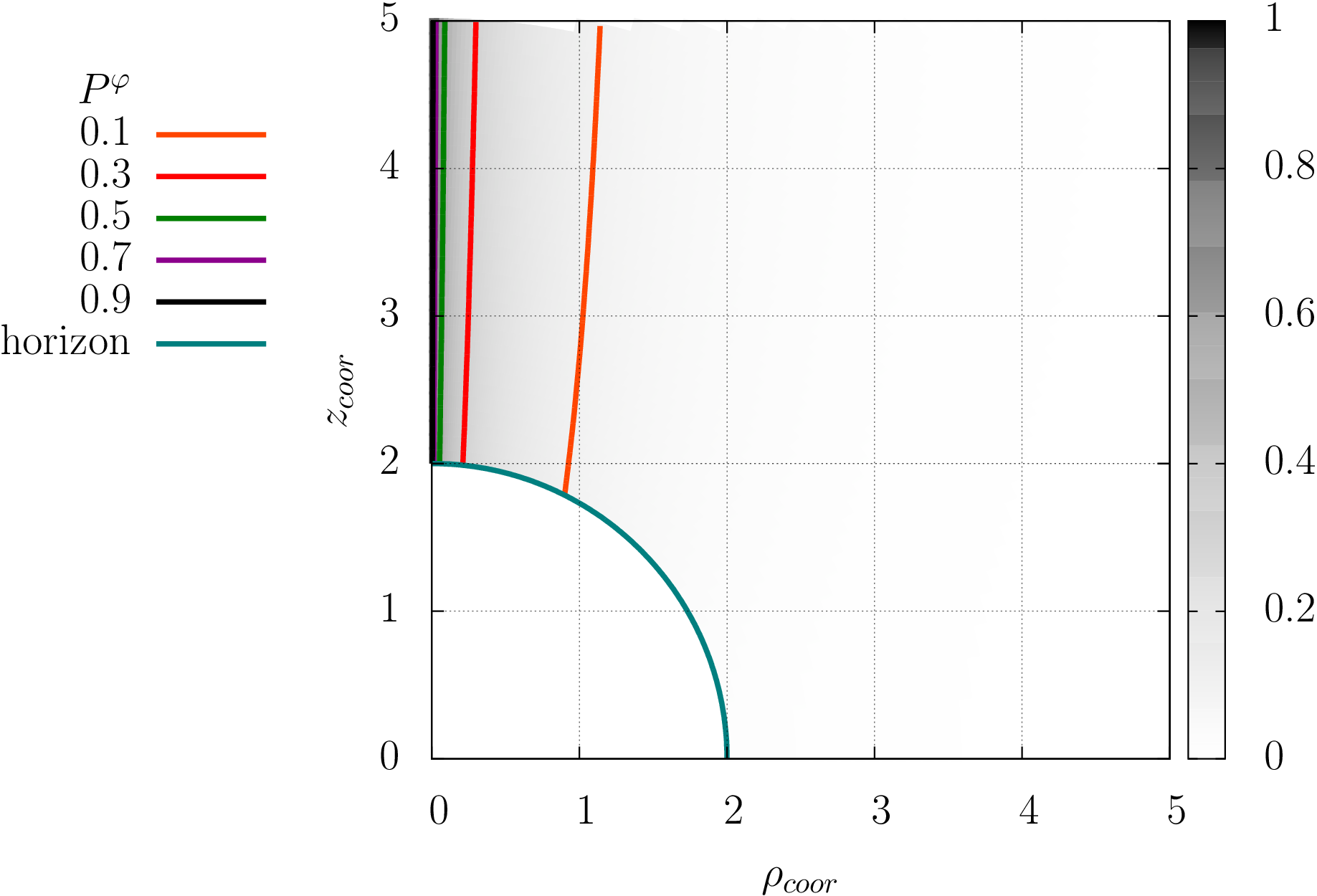}
\label{fig1b}
}\\
\centering
\subfloat[]{
  \includegraphics[width=.46\textwidth]{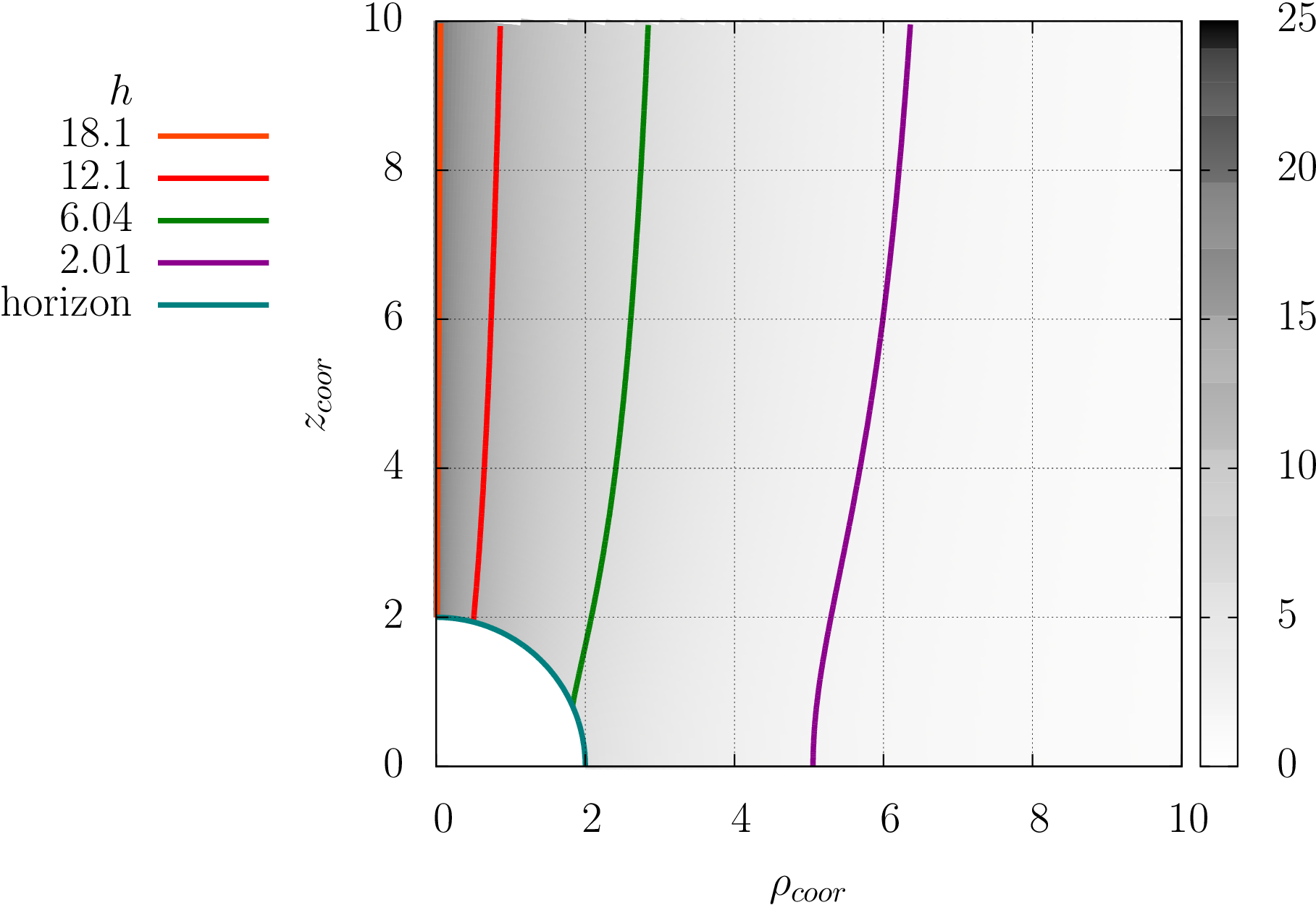}
\label{fig1c}
}
\quad
\subfloat[]{
  \includegraphics[width=.46\textwidth]{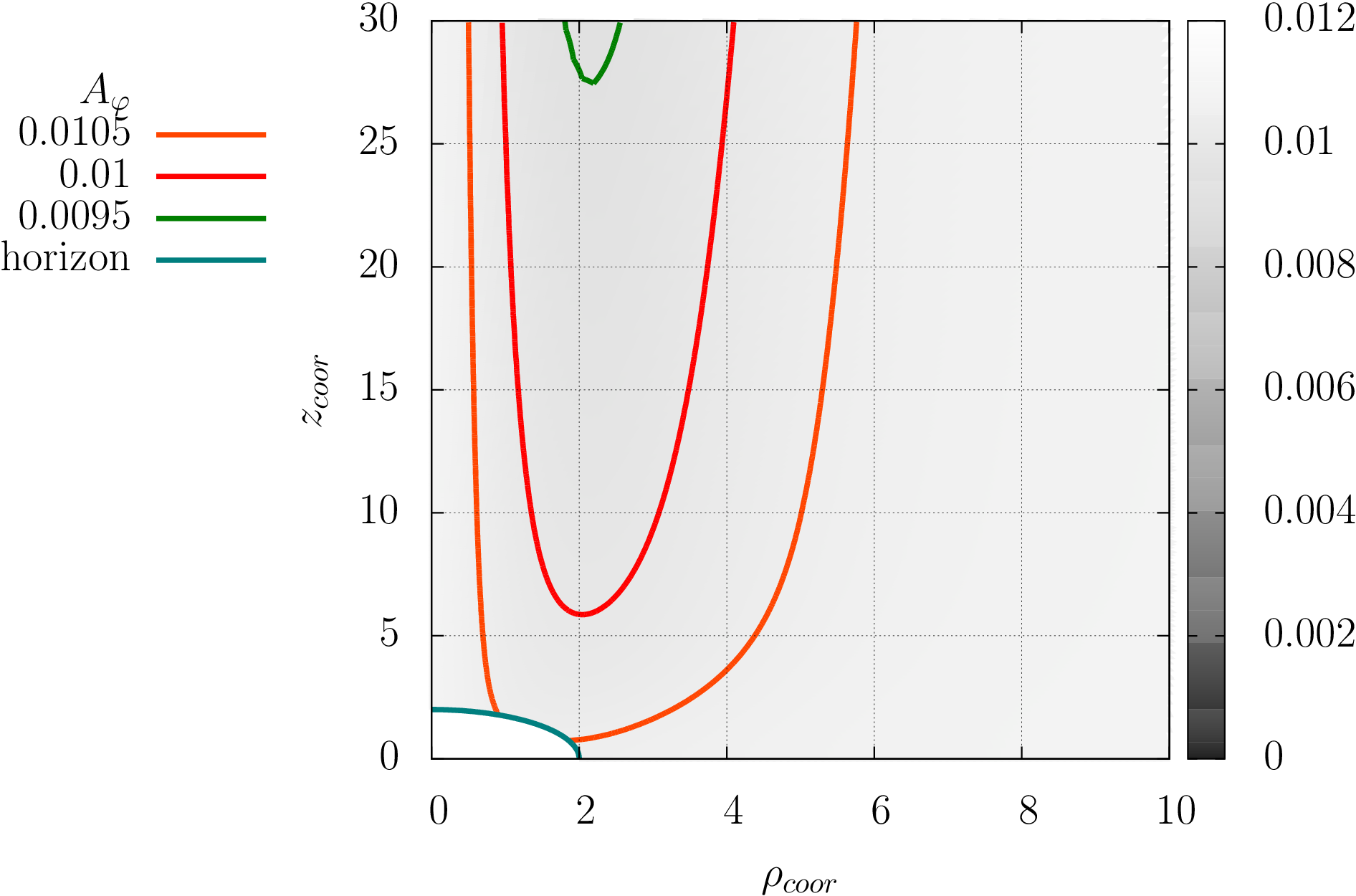}
\label{fig1d}
}
\caption{Cosmic string ($Y,P^{\phi}$), Higgs ($h$) and magnetic ($A_{\phi}$) fields in the Schwarzchild black hole background with $m=0.6$. Axes are $z_{coor} = r \cos{\theta}$,
and $\rho_{coor} = r \sin{\theta}$.}
\label{fig1}
\end{figure}

\begin{figure}[H]
\centering
\subfloat[ ]{
  \includegraphics[width=.46\textwidth]{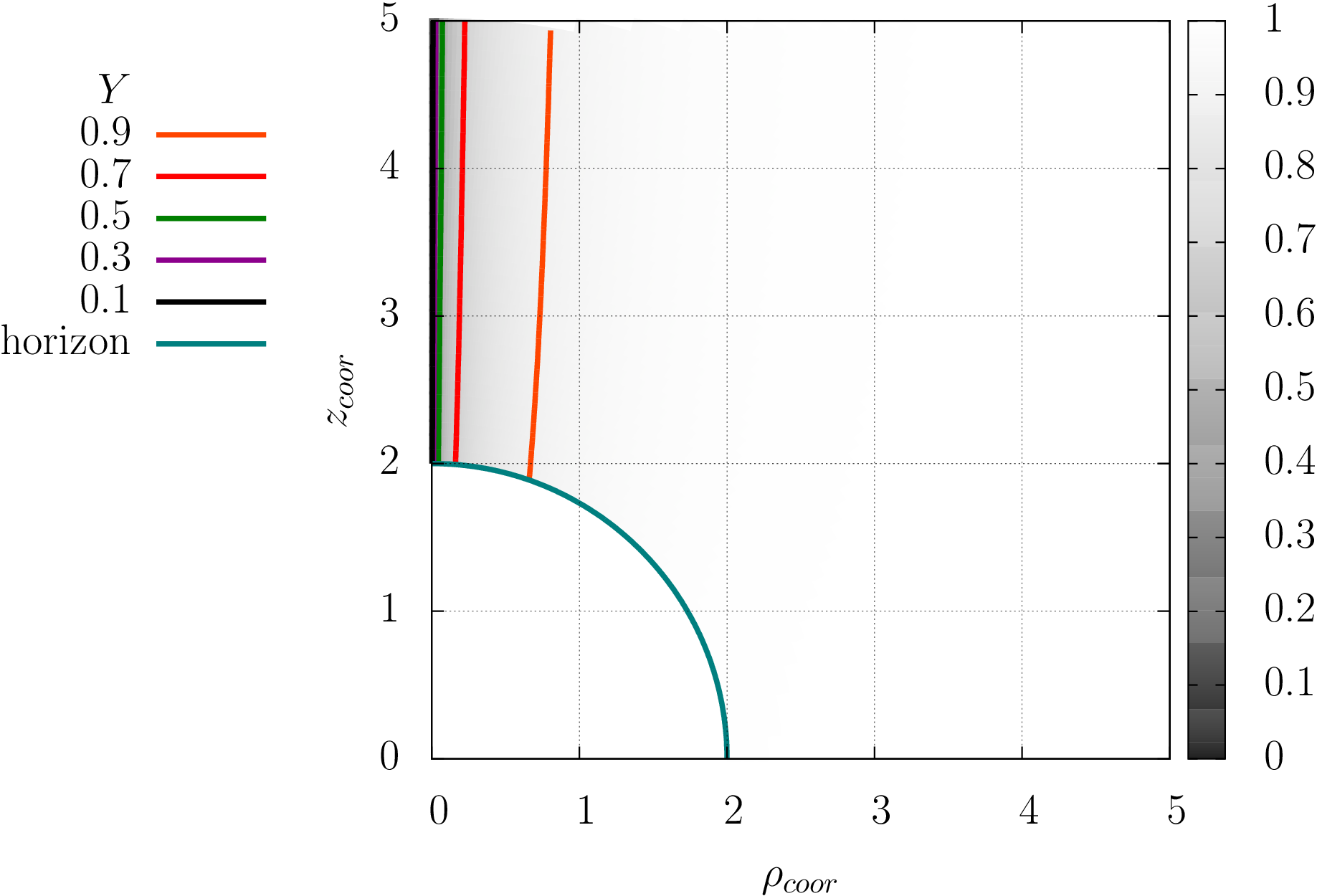}
\label{fig2a}
  }
\quad
\subfloat[]{
  \includegraphics[width=.46\textwidth]{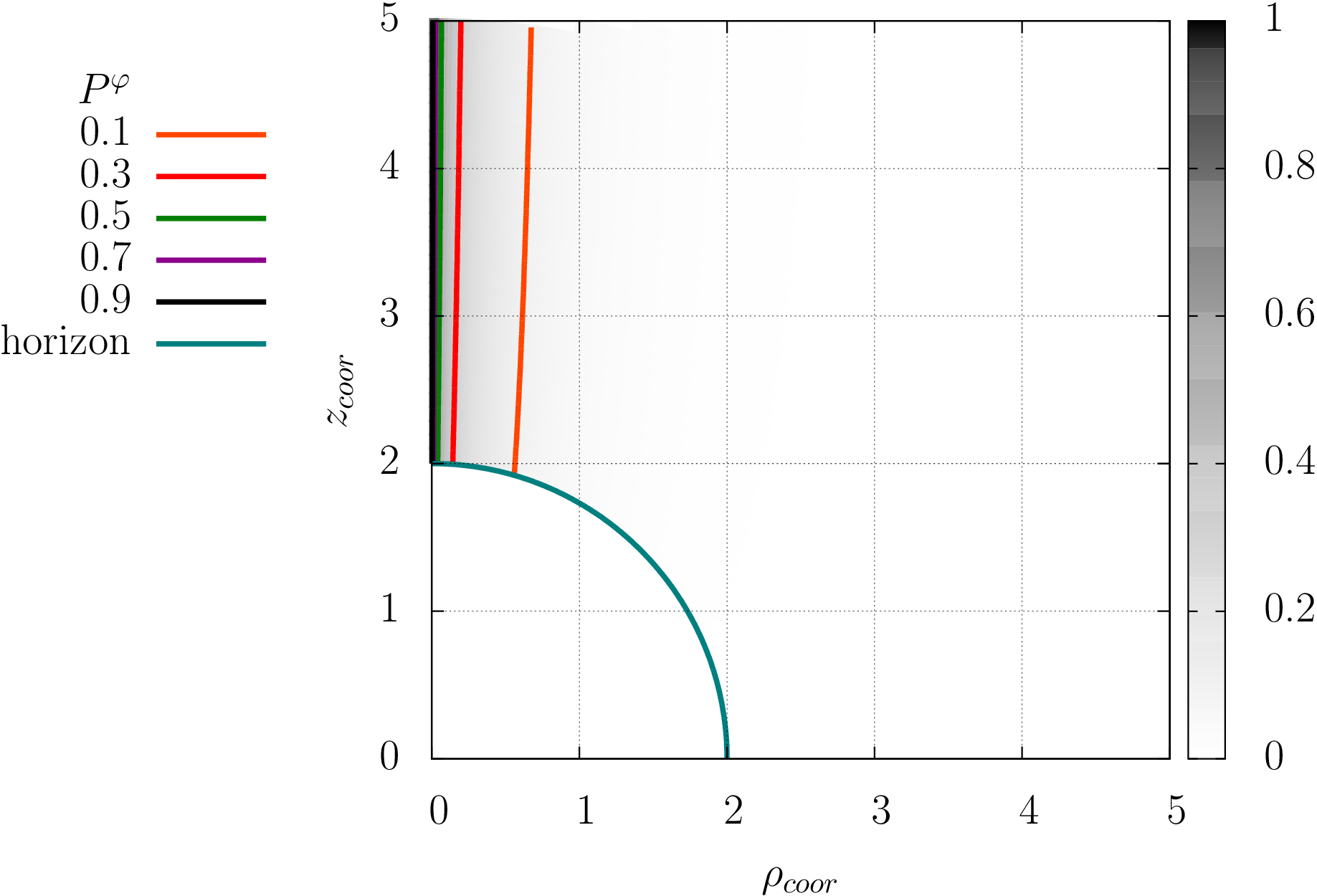}
\label{fig2b}
}\\
\centering
\subfloat[]{
  \includegraphics[width=.46\textwidth]{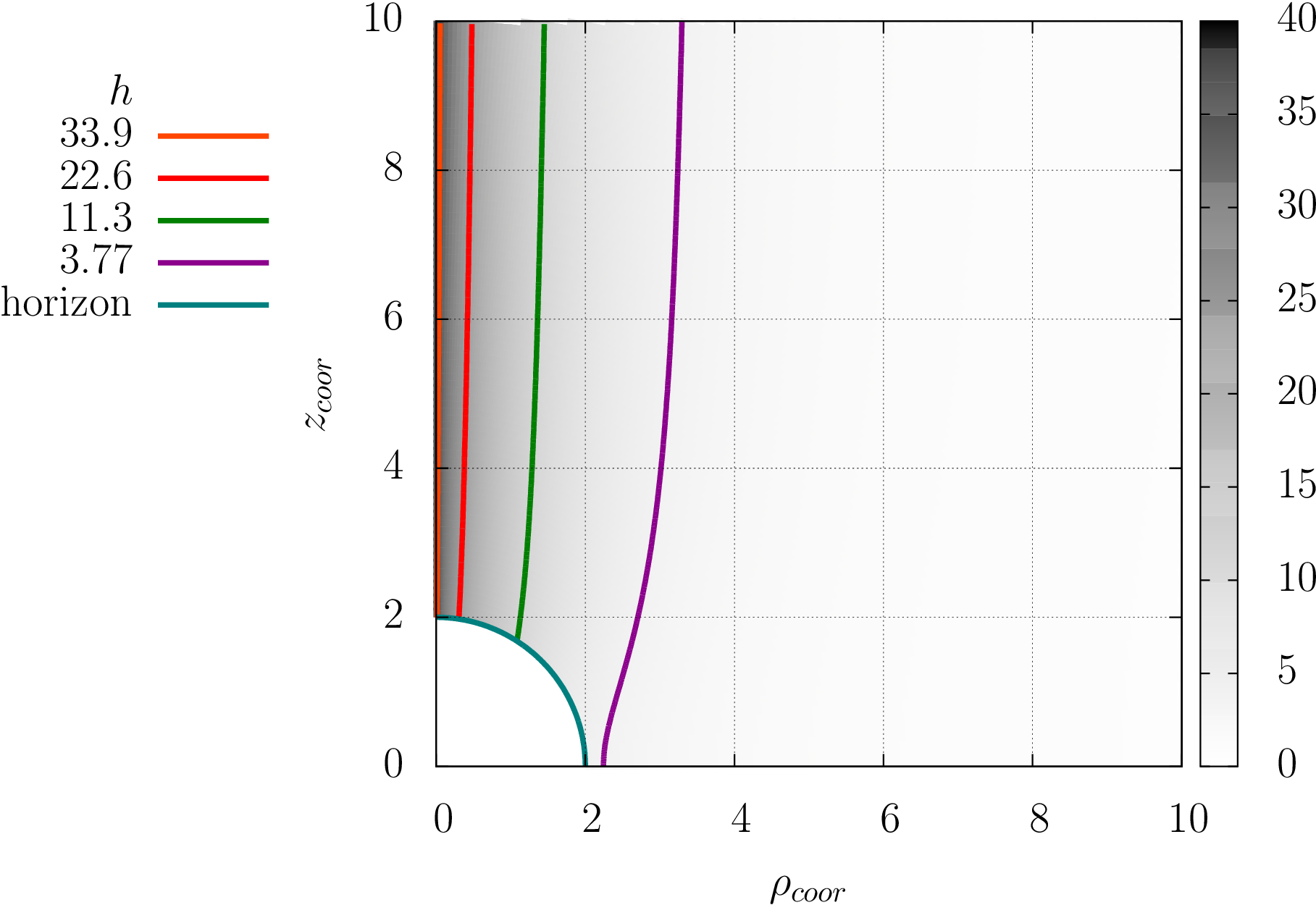}
\label{fig2c}
}
\quad
\subfloat[]{
  \includegraphics[width=.46\textwidth]{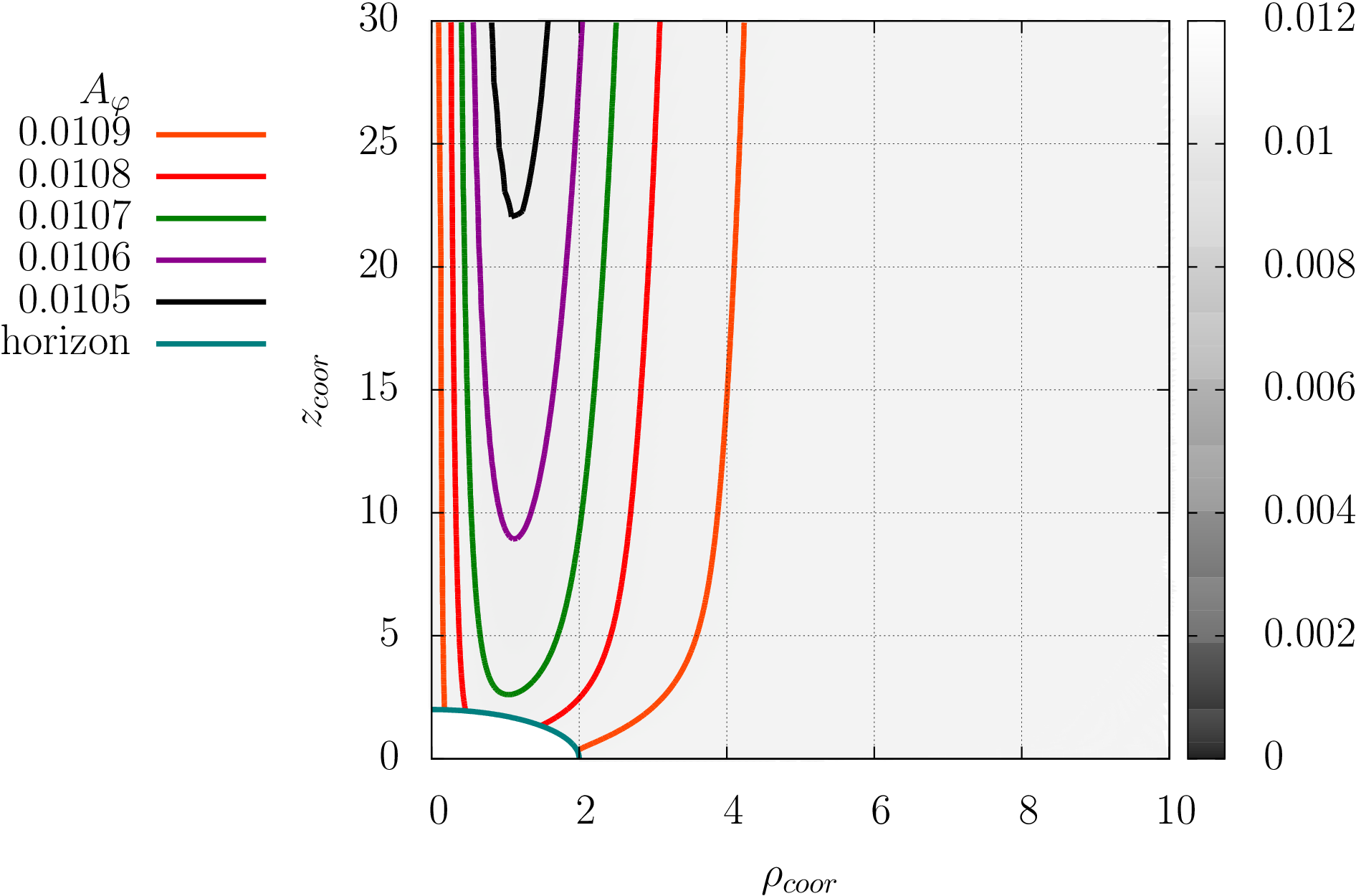}
\label{fig2d}
}
\caption{Cosmic string ($Y,P^{\phi}$), Higgs ($h$) and magnetic ($A_{\phi}$) fields in the Schwarzchild black hole background with $m=2.1$.
Axes are $z_{coor} = r \cos{\theta}$, and $\rho_{coor} = r \sin{\theta}$.}
\label{fig2}
\end{figure}

\begin{figure}[H]
\centering
\subfloat[ ]{
  \includegraphics[width=.46\textwidth]{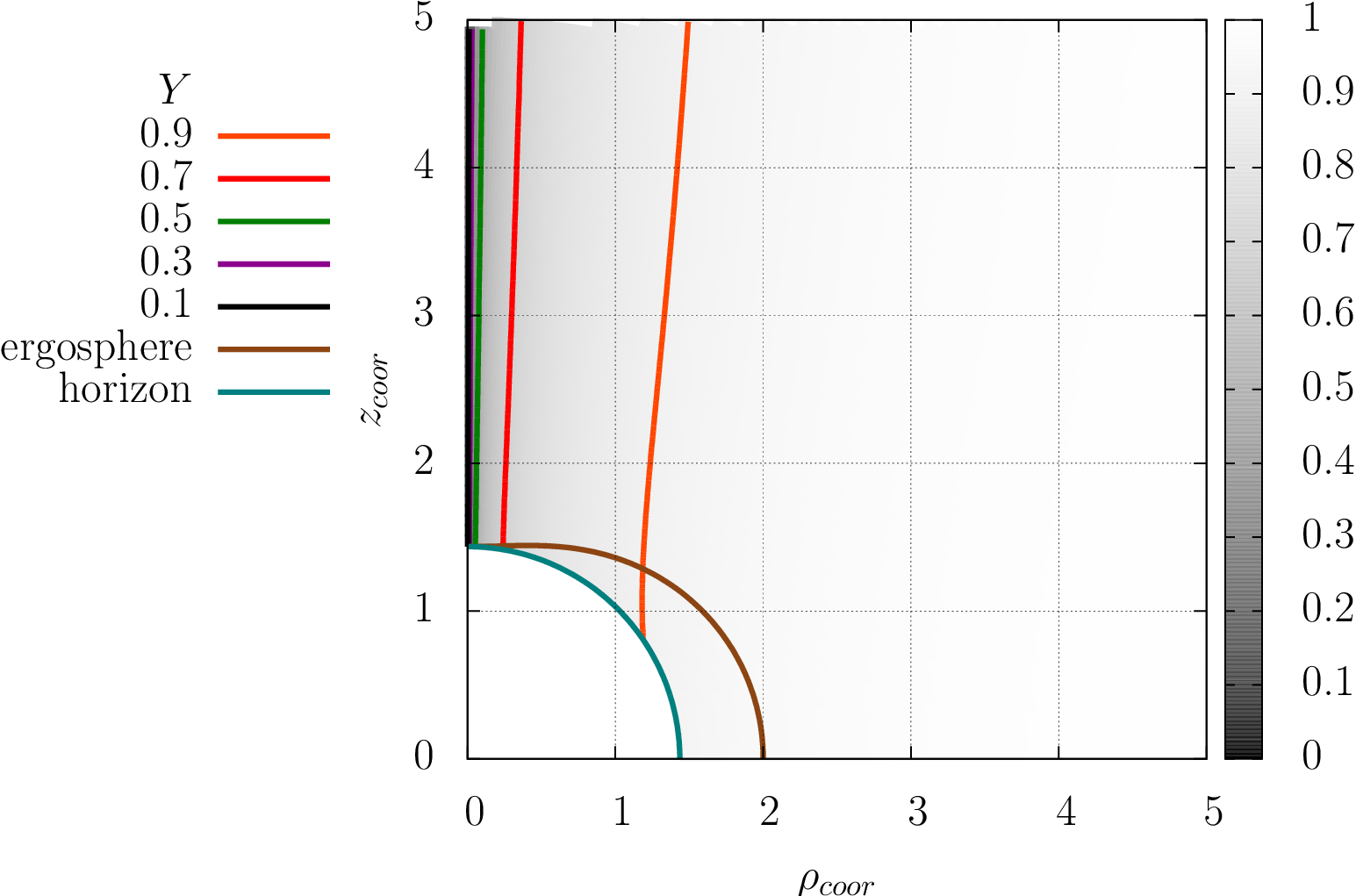}
\label{fig3a}
  }
\quad
\subfloat[]{
  \includegraphics[width=.46\textwidth]{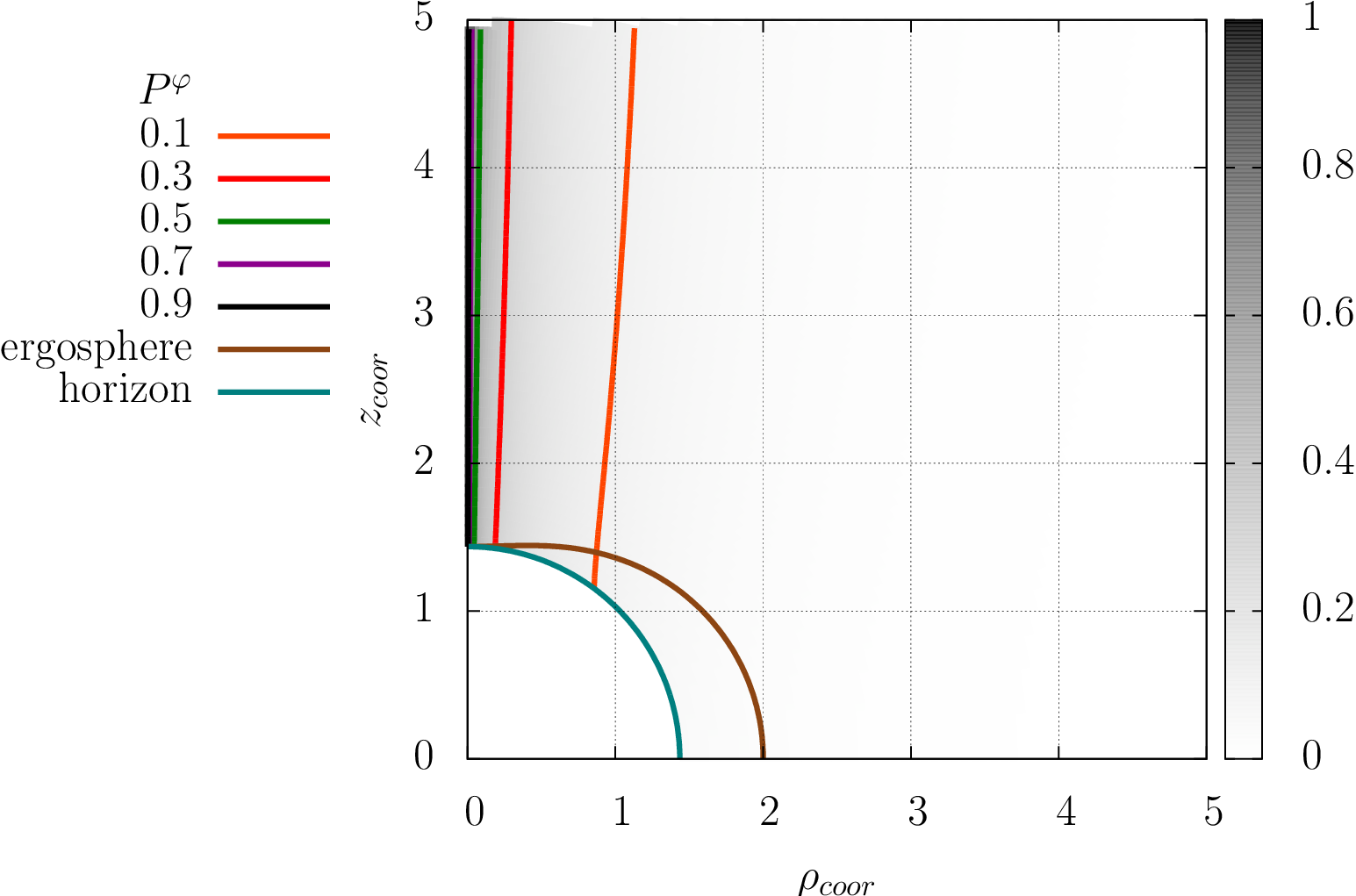}
\label{fig3b}
}\\
\centering
\subfloat[]{
  \includegraphics[width=.46\textwidth]{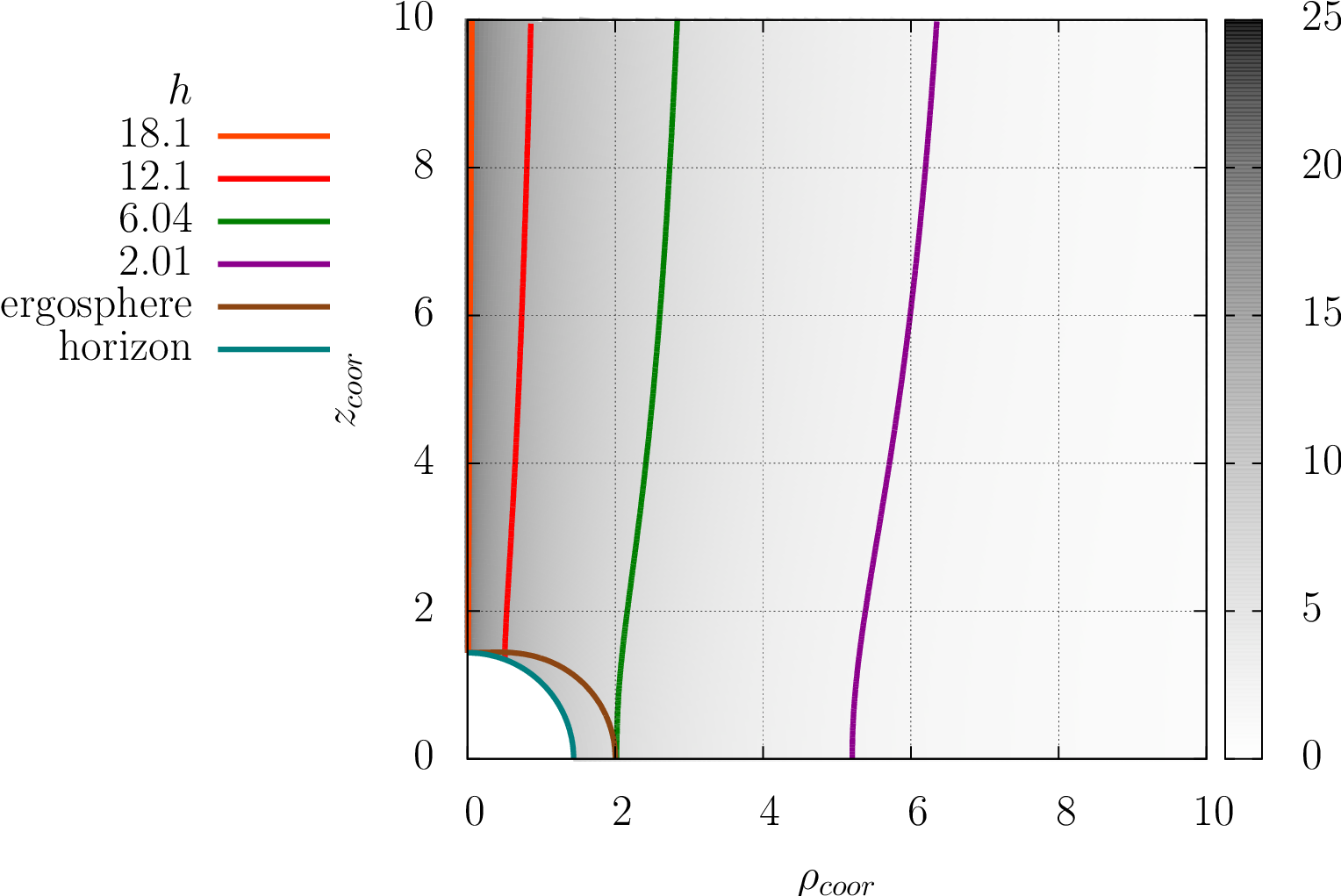}
\label{fig3c}
}
\quad
\subfloat[]{
  \includegraphics[width=.46\textwidth]{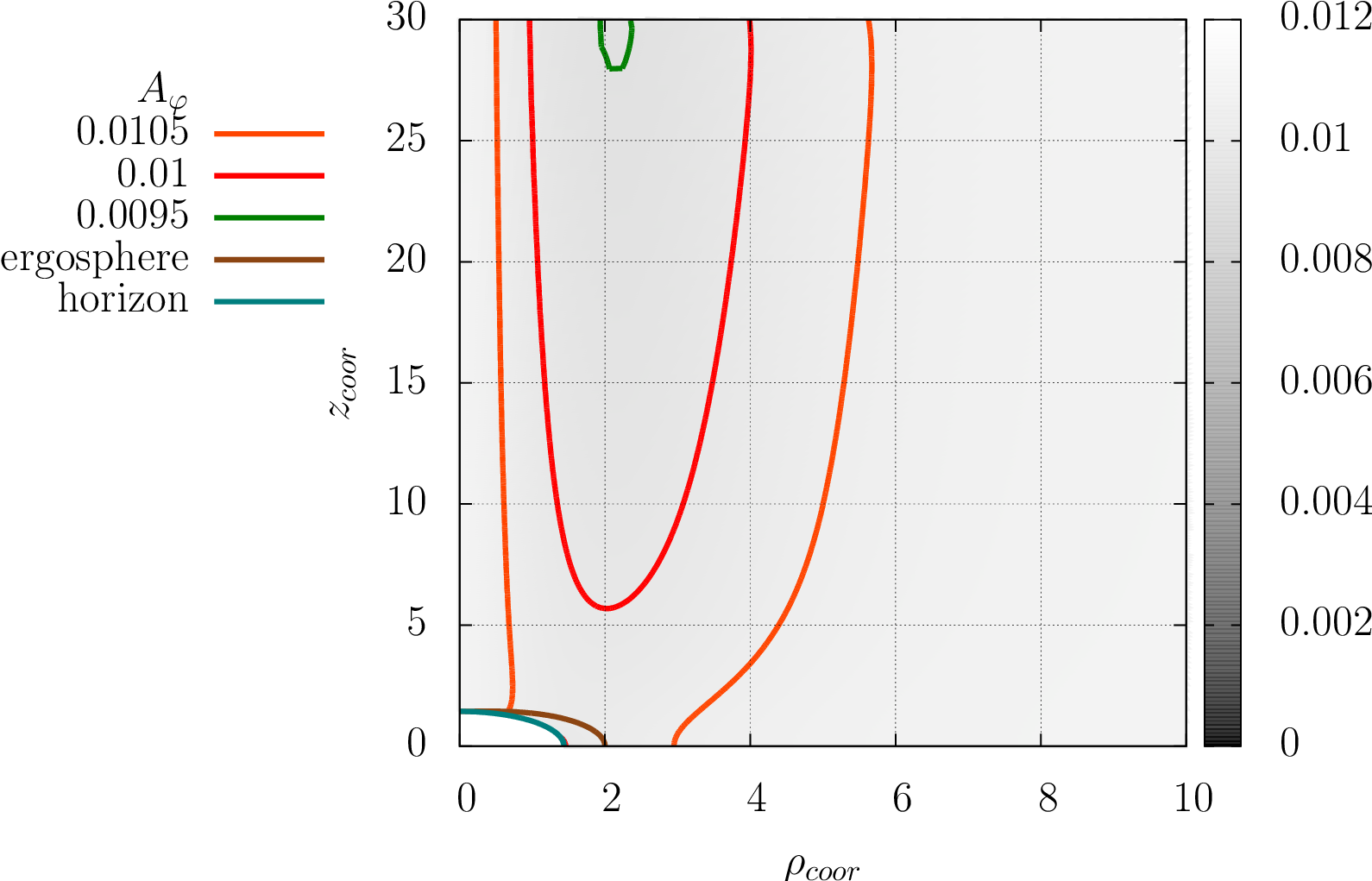}
\label{fig3d}
}
\caption{Cosmic string ($Y,P^{\phi}$), Higgs ($h$) and magnetic ($A_{\phi}$) fields in the Kerr black hole spacetime with $m=0.6$ and $a=0.9$.
Axes are $z_{coor} = r \cos{\theta}$, and $\rho_{coor} = r \sin{\theta}$.}
\label{fig3}
\end{figure}

\begin{figure}[H]
\centering
\subfloat[ ]{
  \includegraphics[width=.46\textwidth]{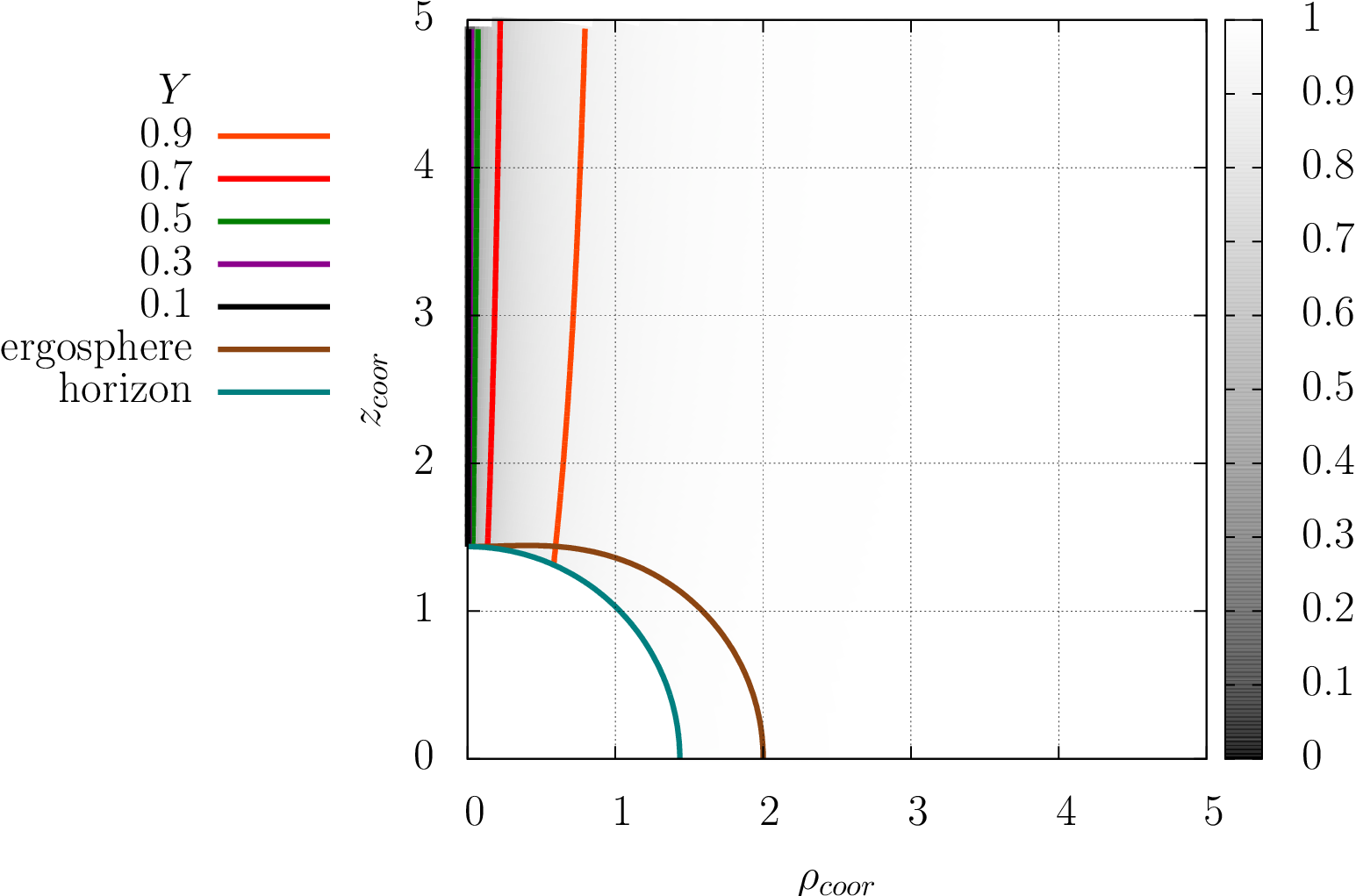}
\label{fig4a}
  }
\quad
\subfloat[]{
  \includegraphics[width=.46\textwidth]{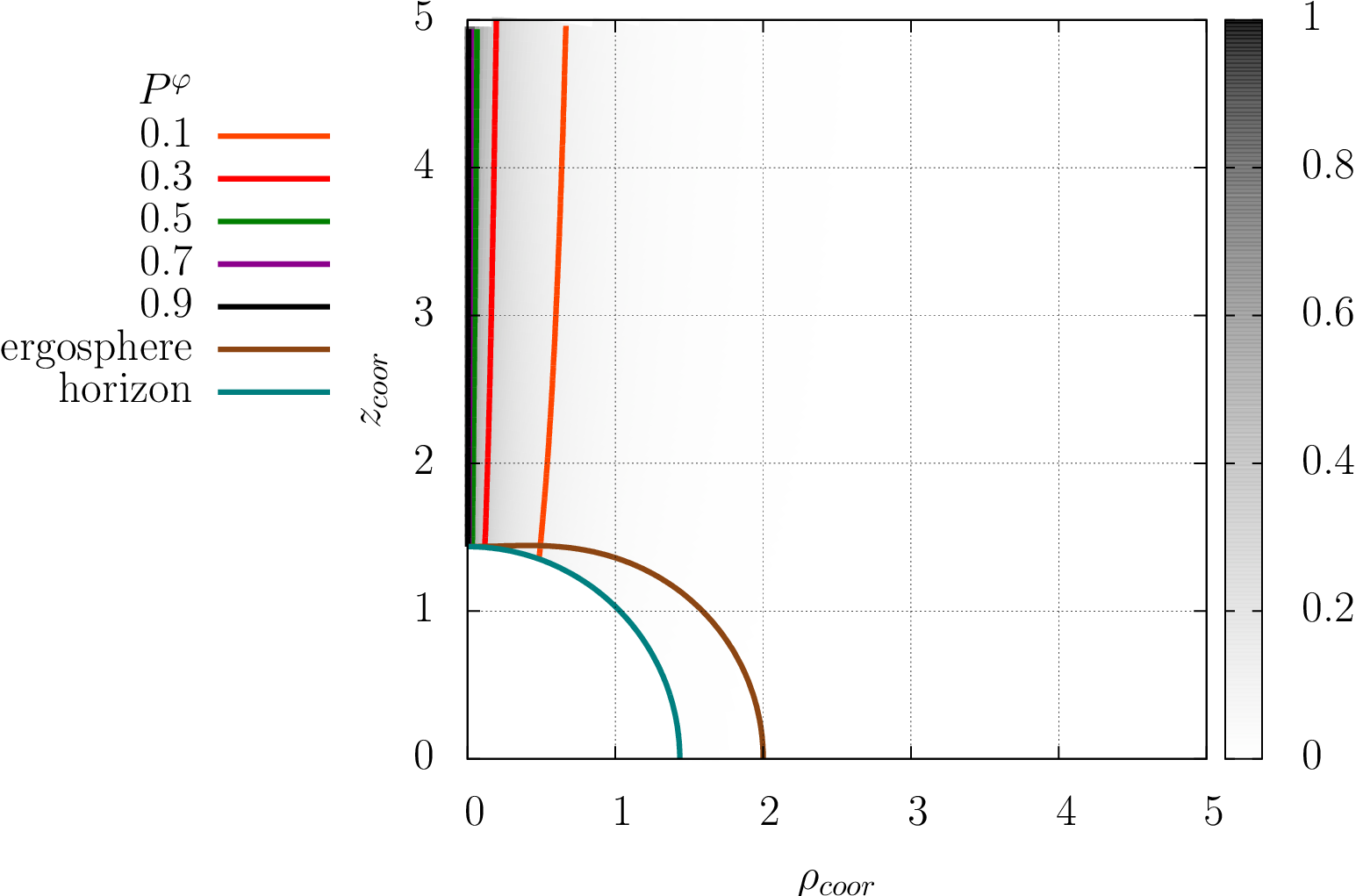}
\label{fig4b}
}\\
\centering
\subfloat[]{
  \includegraphics[width=.46\textwidth]{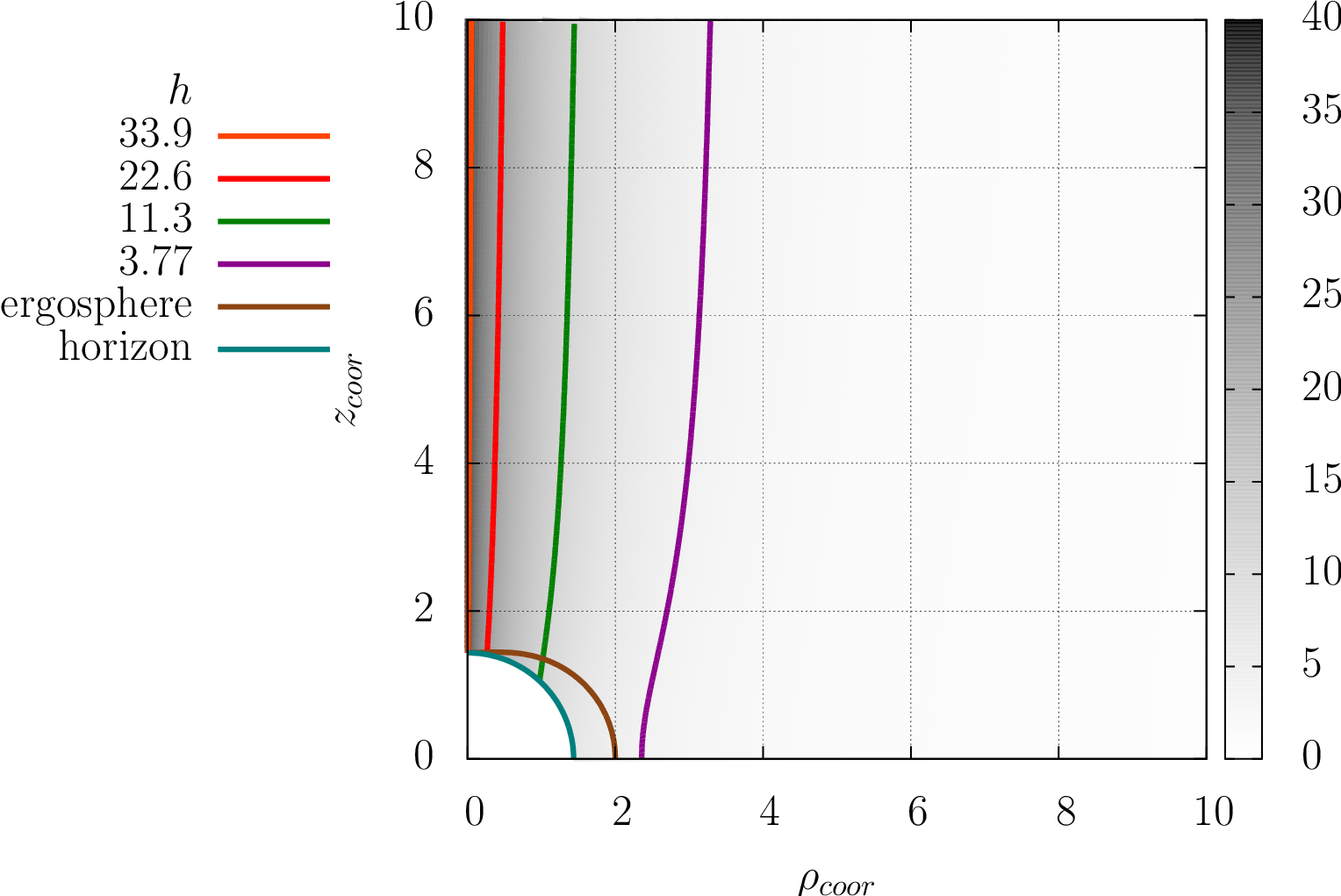}
\label{fig4c}
}
\quad
\subfloat[]{
  \includegraphics[width=.46\textwidth]{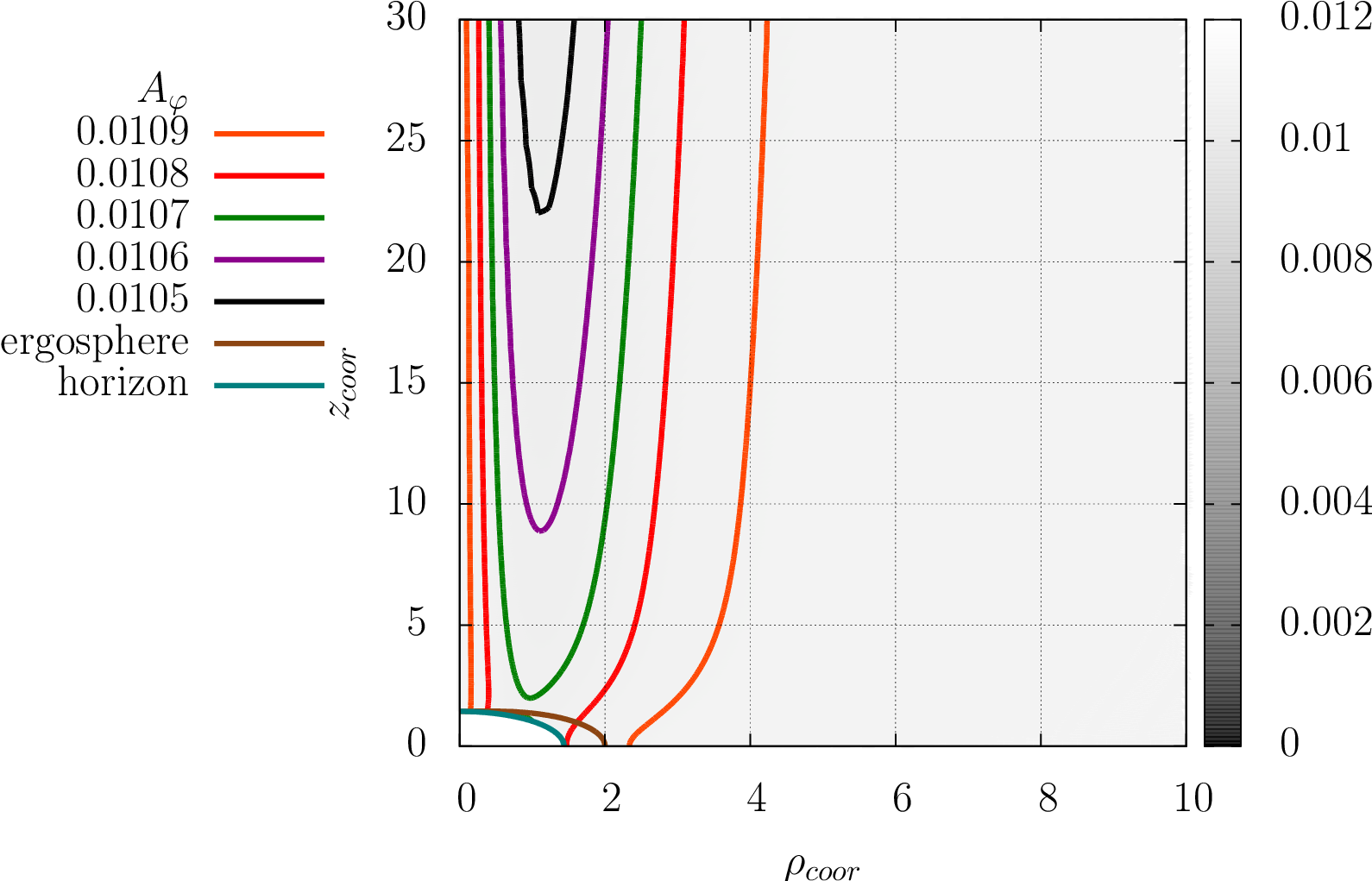}
\label{fig4d}
}
\caption{Cosmic string ($Y,P^{\phi}$), Higgs ($h$) and magnetic ($A_{\phi}$) fields in the Kerr black hole spacetime with $m=2.1$ and $a=0.9$.
Axes are $z_{coor} = r \cos{\theta}$, and $\rho_{coor} = r \sin{\theta}$.}
\label{fig4}
\end{figure}

\begin{figure}[H]
\centering
\subfloat[]{
  \includegraphics[width=.46\textwidth]{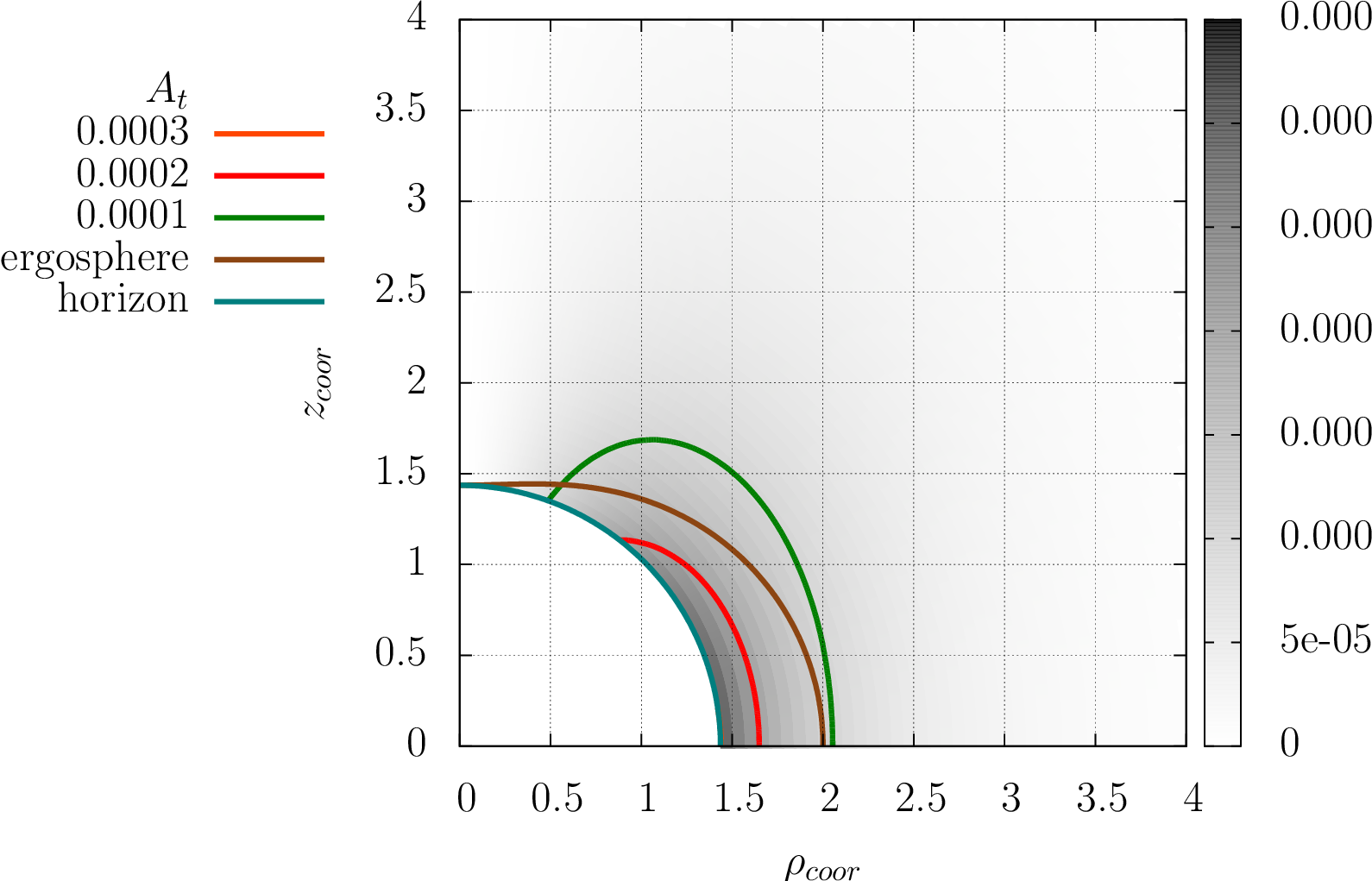}
\label{fig5a}
}
\quad
\subfloat[]{
  \includegraphics[width=.46\textwidth]{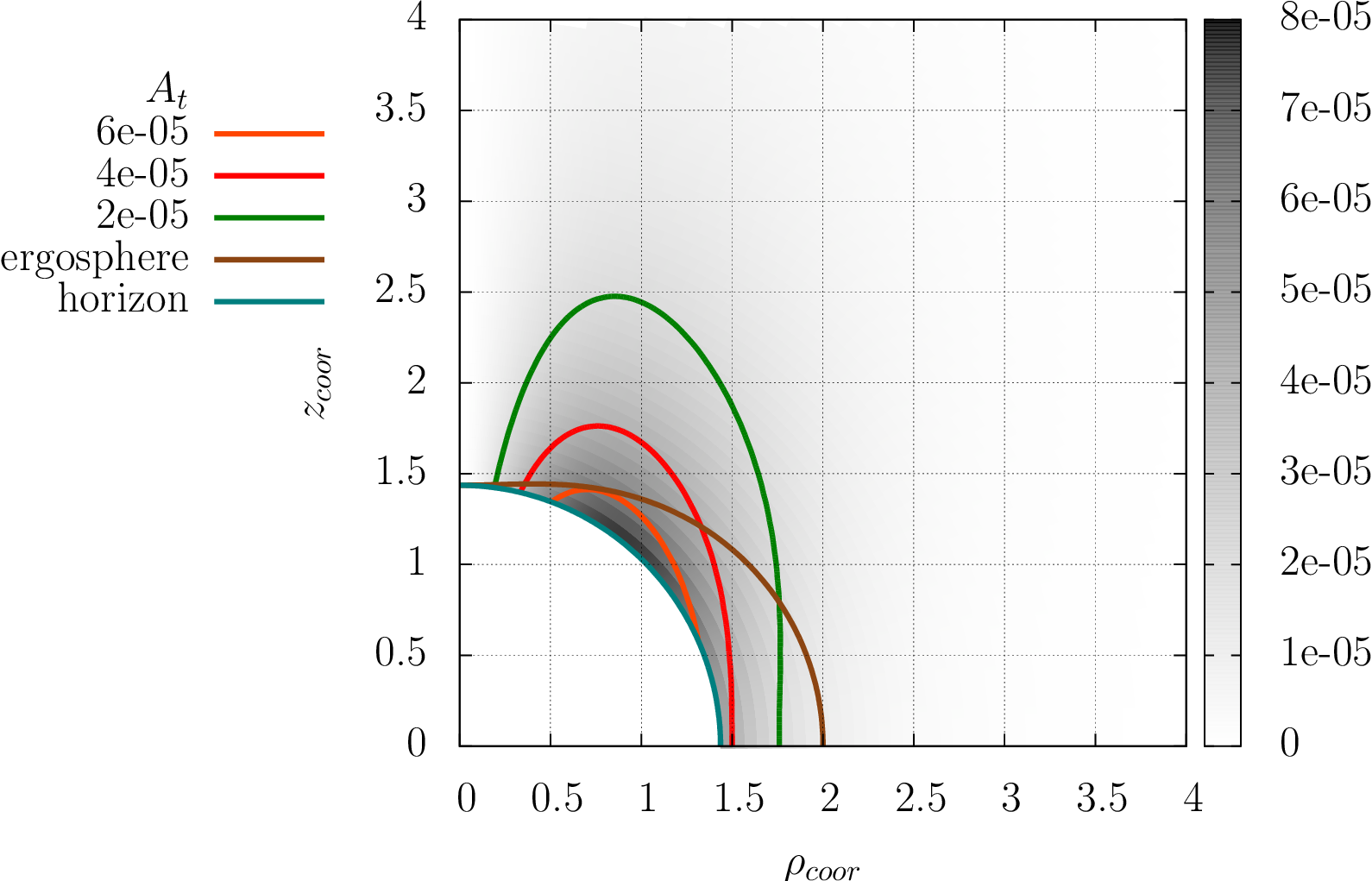}
\label{fig5b}
}
\caption{Time component of the Maxwell field four-potential ($A_{t}$) in the Kerr black hole spacetime with $m=0.6$ (left) and $m=2.1$ (right).
Axes are $z_{coor} = r \cos{\theta}$, and $\rho_{coor} = r \sin{\theta}$.}
\label{fig5}
\end{figure}

Figures \ref{fig1} to \ref{fig5} depict the results of the numerical solution of the system of the equations~(\ref{eomX})-(\ref{eomC}) 
in the case when the gravitational background is given by the 
Schwarzchild and nonextremal Kerr black holes. Comparing the figures \ref{fig1a} and \ref{fig1c}  one can notice
that the presence of the cosmic string leaves its imprint in the Higgs field configuration.
Namely, there is a spatial variation in it that mirrors the $Y$ field distribution. From figure \ref{fig1c} we may infer that the $h$ field achieves 
its maximum value on the symmetry axis (where by definition $Y=0$) and then decreases towards $h_{nostring}$ given in (\ref{fieldsNoString}). 
Interpreting this result in the language of the physical Higgs field $H_{-}$ (the field that is given by the eigenstate of the mass matrix (\ref{M2f})), 
we conclude that far away from the black hole this field is a mixture of $h$ and $Y$, while when one approaches the symmetry axis,
the admixture of the $Y$ field is decreasing and eventually on the symmetry axis $H_{-}$ is composed entirely of the $h$ field. 
Moreover, we see that the region where we have a nonzero gradient of the $h$ field is wider than the cosmic string core (i.e., the region where there is a nonzero $Y$ field gradient). 
This result is in agreement with the flat spacetime results obtained in~\cite{hyd14}, where it was also noticed that the Higgs field forms a wider cloud around the cosmic string. 
At this point it is interesting to remark that a similar structure was observed in the case of the cosmic string-fermions-dark string system 
in~\cite{nak12} where the authors found that for some values of the parameters the fermions form a layer around the cosmic string core.

From figures \ref{fig2a} and \ref{fig2c} we may see that the bigger the $m$ parameter is, the smaller the cosmic string core is (and $h$ field cloud surrounding it also shrinks). 
The cause of this becomes obvious if we remember that in our setup the dimensionless parameter $m$ contains a product of the black hole mass $M$ and 
the mass parameter $m_{Y}$ of the $Y$ field. Since by the definition we keep $m_{Y}$ fixed, the change of $m$ is 
equivalent to the change of the black hole mass which in turn due to our choice of the rescaling is reflected in the change of the string width.

After the analysis of the scalar sector of the considered system, let us examine its gauge sector. In figure \ref{fig1b} we see the typical configuration of the magnetic component of the gauge field 
forming the cosmic string, namely the field is nonzero in the string core and decays to zero outside it. On the other hand, from 
the figure 
\ref{fig1d} we may infer that in the considered spacetime there is a nontrivial configuration
of the electromagnetic field. It is caused due to
the kinetic coupling between the electromagnetic and {\it dark photon} gauge fields. In the case of the Schwarzchild black hole the presence of the cosmic string induces only
a nontrivial magnetic field, while for the Kerr black hole also the electric field is induced. The last fact is due to the presence of the off-diagonal components of the metric tensor in the case of the 
rotating black hole. The temporal component of the electromagnetic four potential that leads to the nontrivial electric field is depicted in figure \ref{fig5}. Beside the presence of the nonzero $A_t$ component 
in the case of the Kerr background there is no visible difference in the fields configuration for the Schwarzschild and nonextremal Kerr black holes.

\begin{figure}[H]
\centering
\subfloat[ ]{
  \includegraphics[width=.46\textwidth]{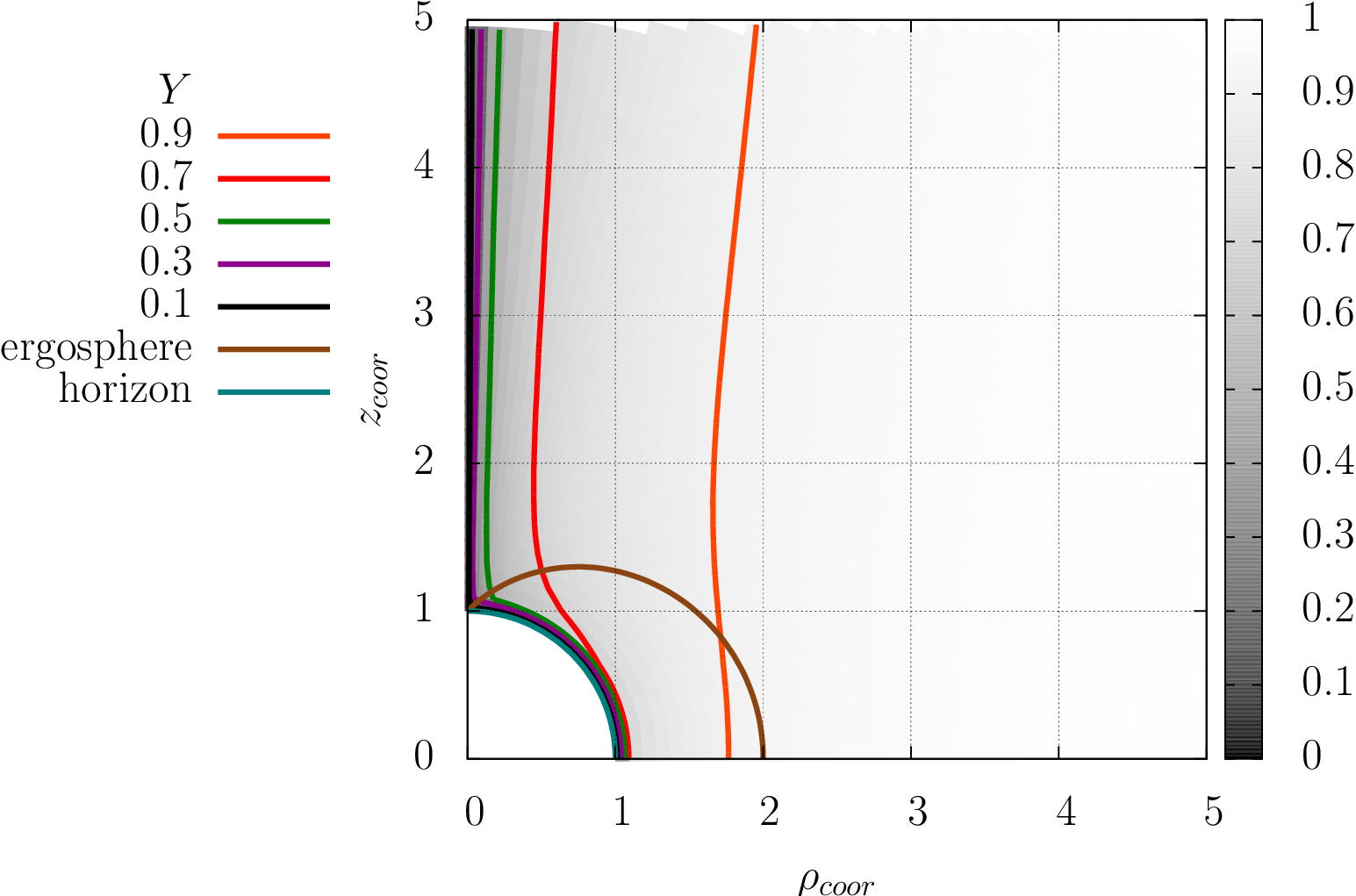}
\label{fig6a}
  }
\quad
\subfloat[]{
  \includegraphics[width=.46\textwidth]{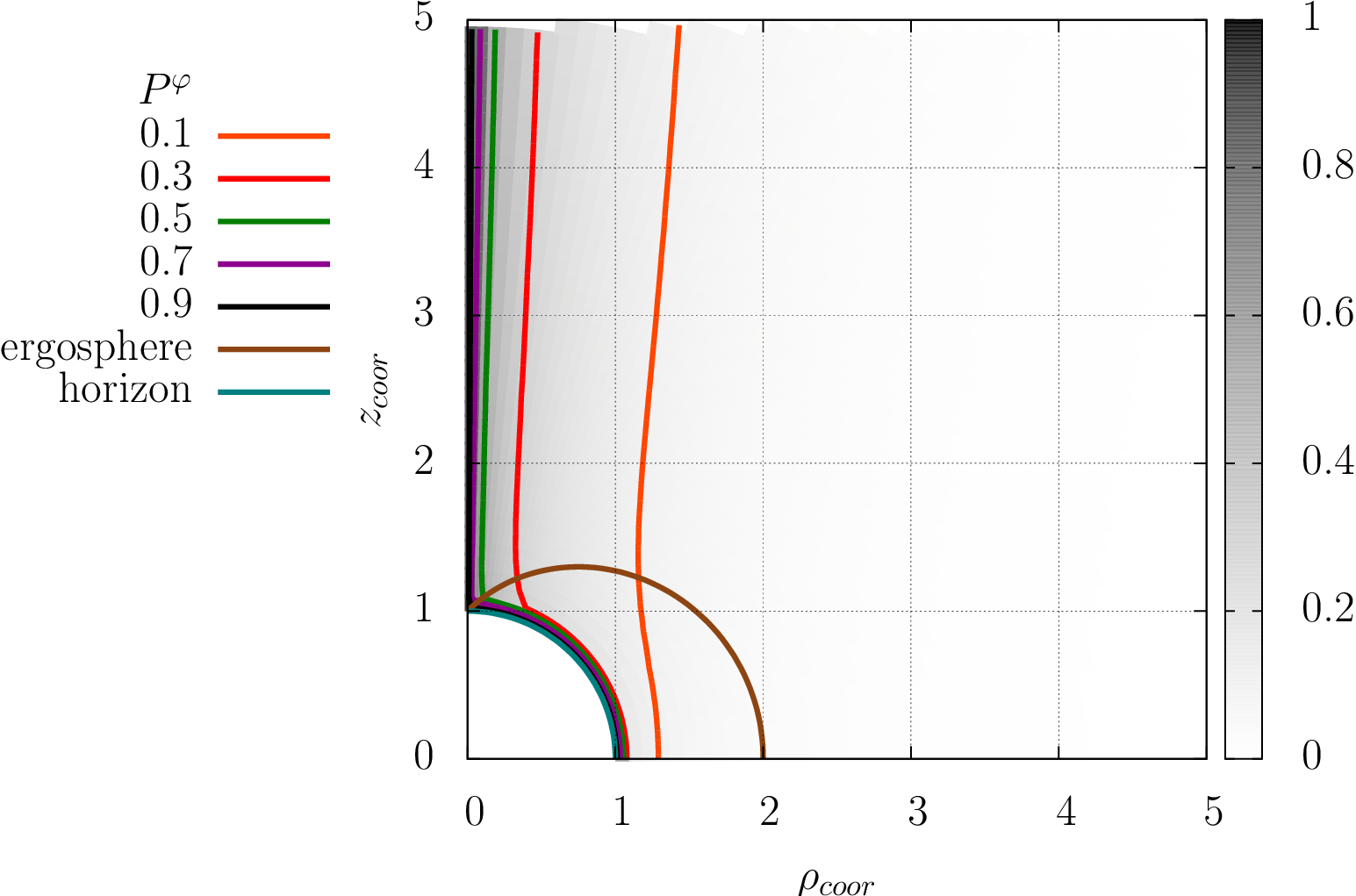}
\label{fig6b}
}\\
\centering
\subfloat[]{
  \includegraphics[width=.46\textwidth]{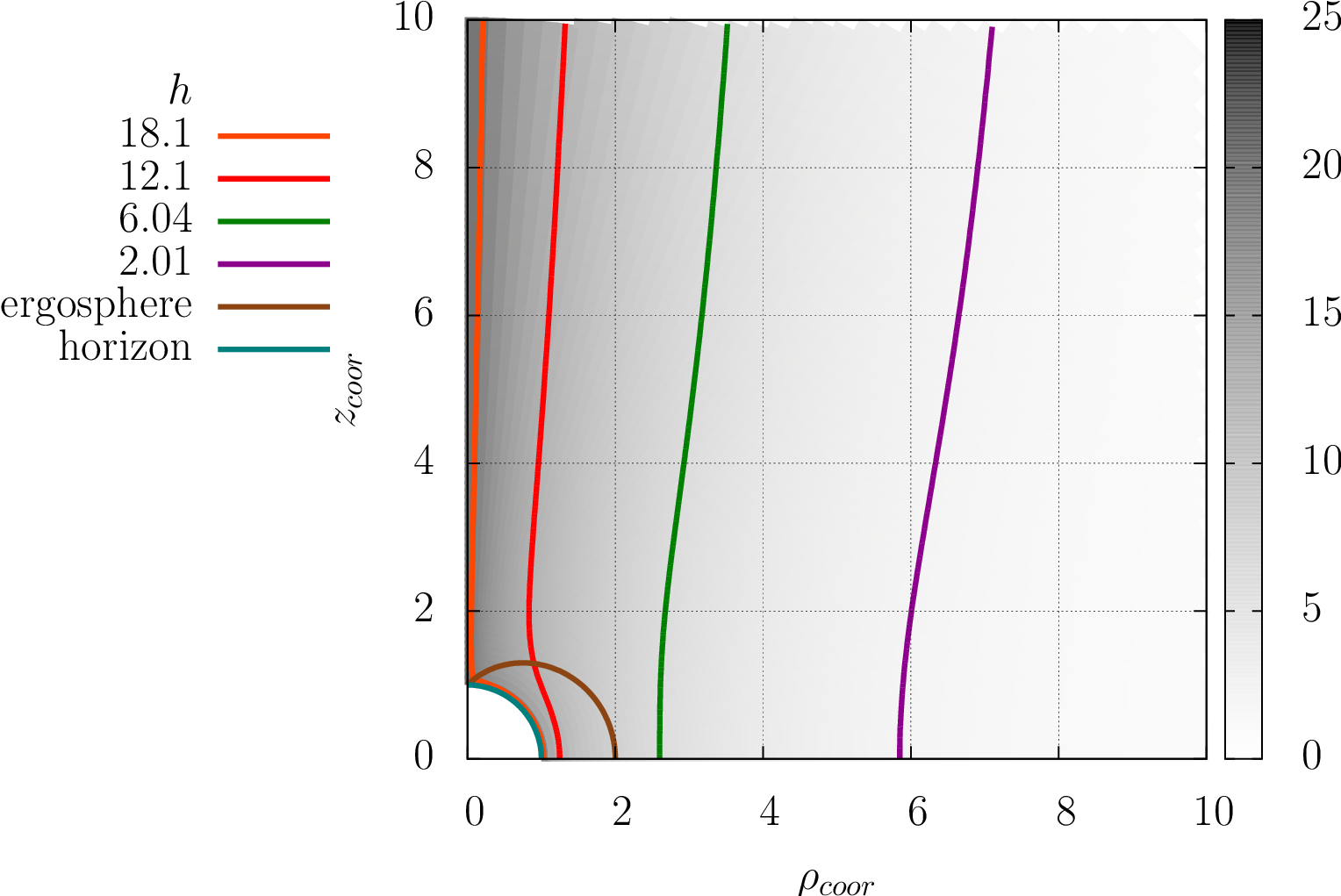}
\label{fig6c}
}
\caption{Cosmic string ($Y,P^{\phi}$) and Higgs ($h$) fields in the extremal Kerr black hole spacetime with $m=0.6$ and $a=1.0$.
Axes are $z_{coor} = r \cos{\theta}$, and $\rho_{coor} = r \sin{\theta}$.}
\label{fig6}
\end{figure}

\begin{figure}[H]
\centering
\subfloat[ ]{
  \includegraphics[width=.46\textwidth]{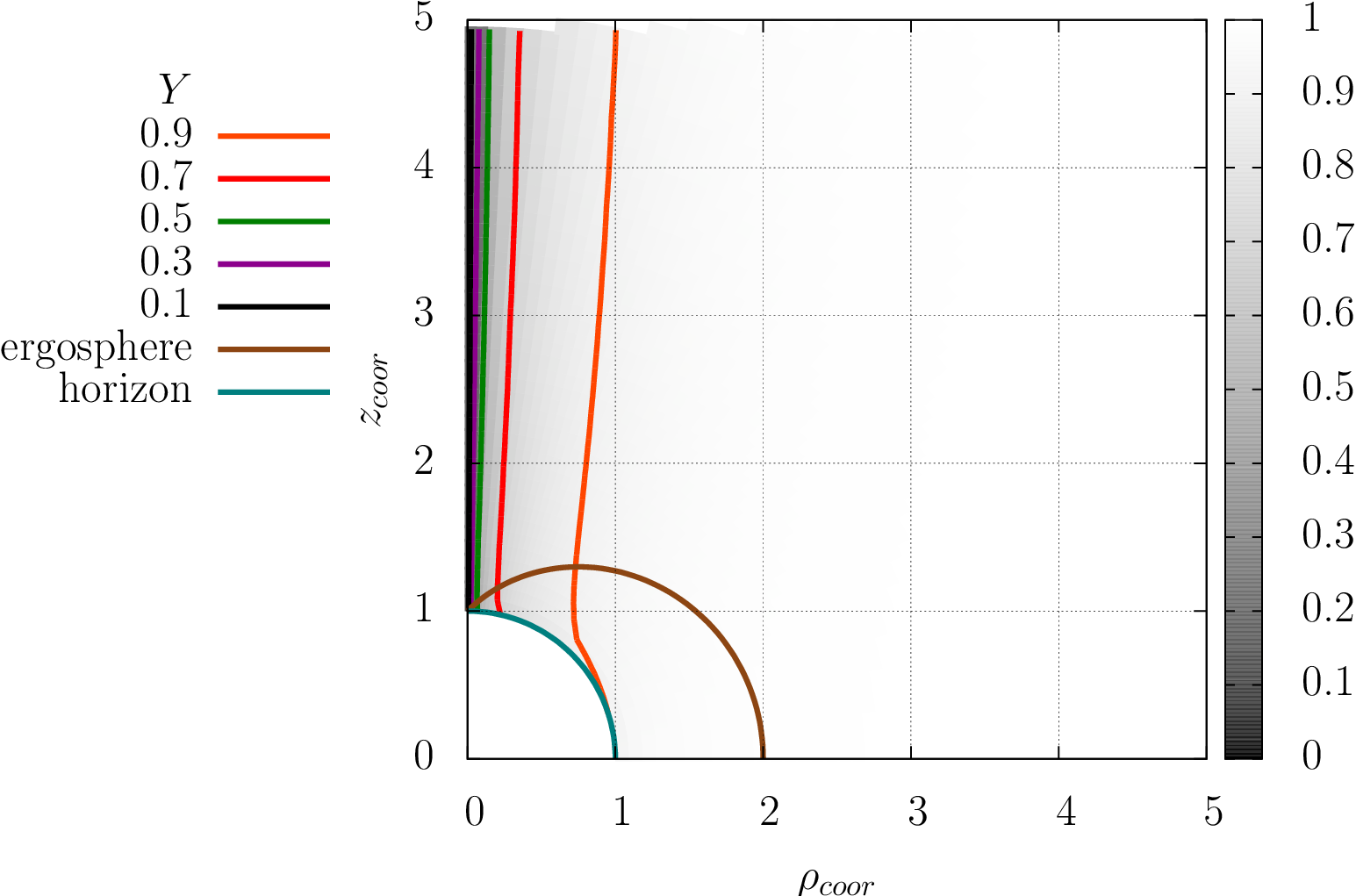}
\label{fig7a}
  }
\quad
\subfloat[]{
  \includegraphics[width=.46\textwidth]{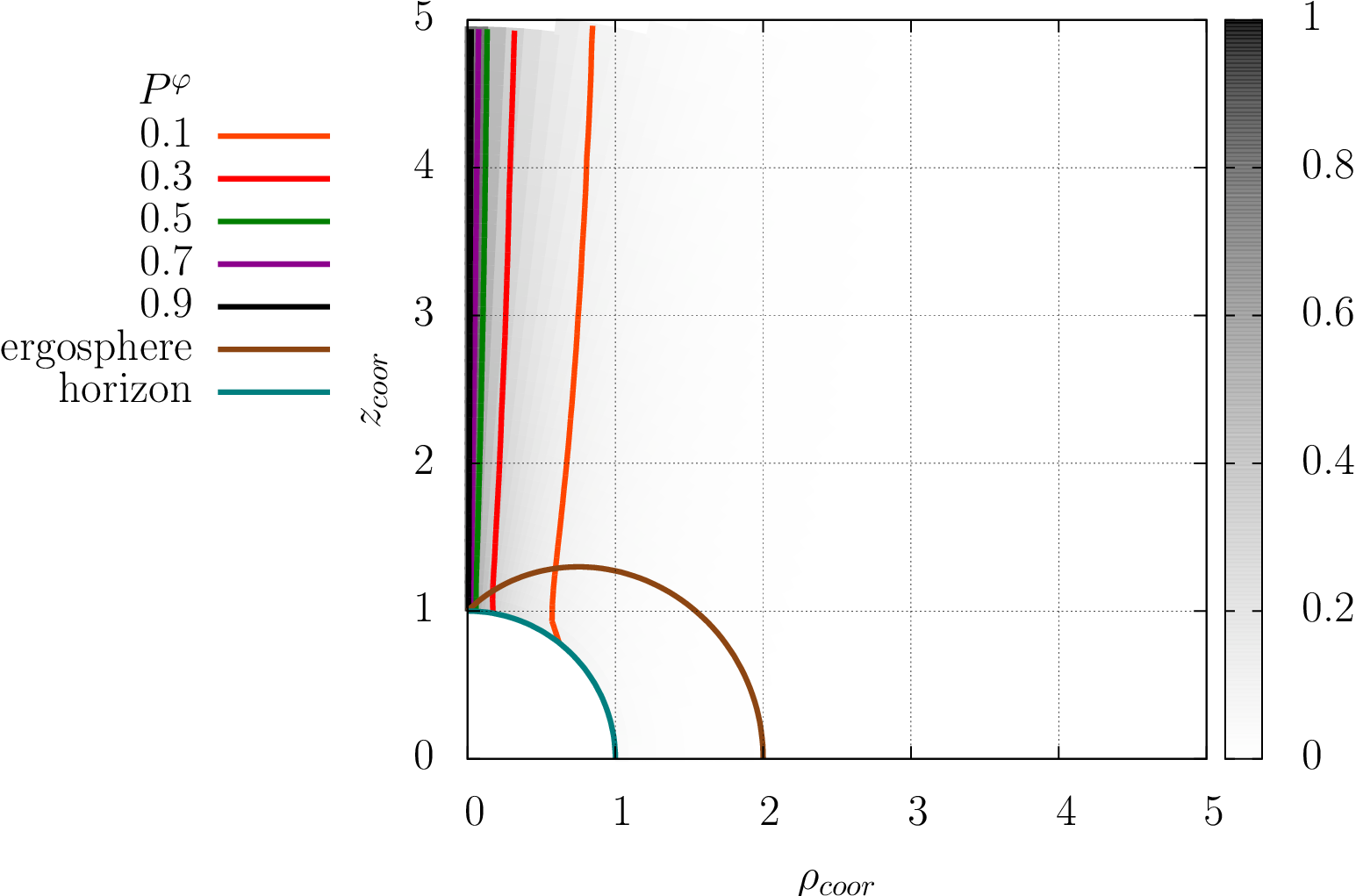}
\label{fig7b}
}\\
\centering
\subfloat[]{
  \includegraphics[width=.46\textwidth]{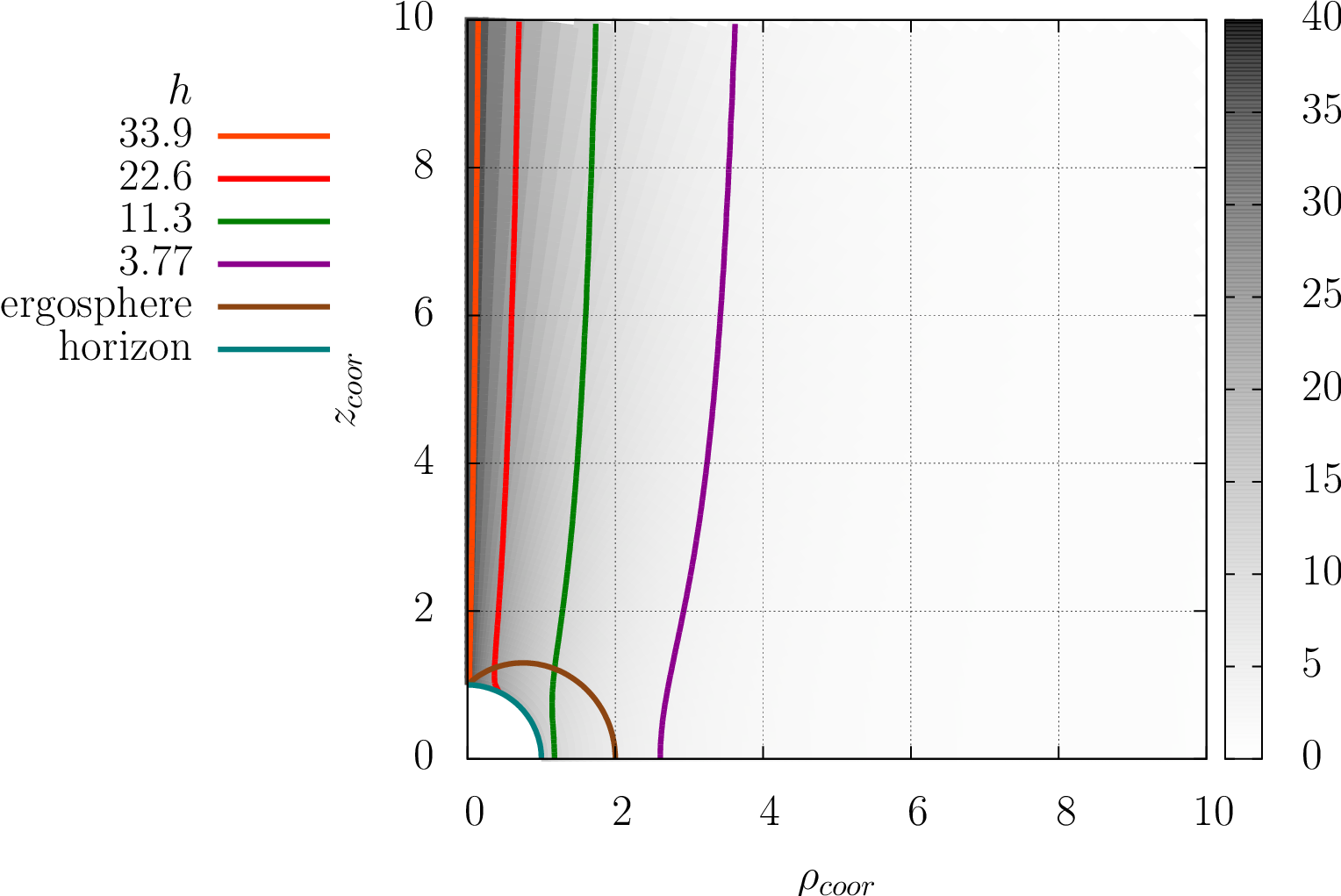}
\label{fig7c}
}
\caption{Cosmic string ($Y,P^{\phi}$) and Higgs ($h$) fields in the extremal Kerr black hole spacetime with $m=2.1$ and $a=1.0$.
Axes are $z_{coor} = r \cos{\theta}$, and $\rho_{coor} = r \sin{\theta}$.}
\label{fig7}
\end{figure}

\begin{figure}[H]
\centering
\subfloat[ ]{
  \includegraphics[width=.46\textwidth]{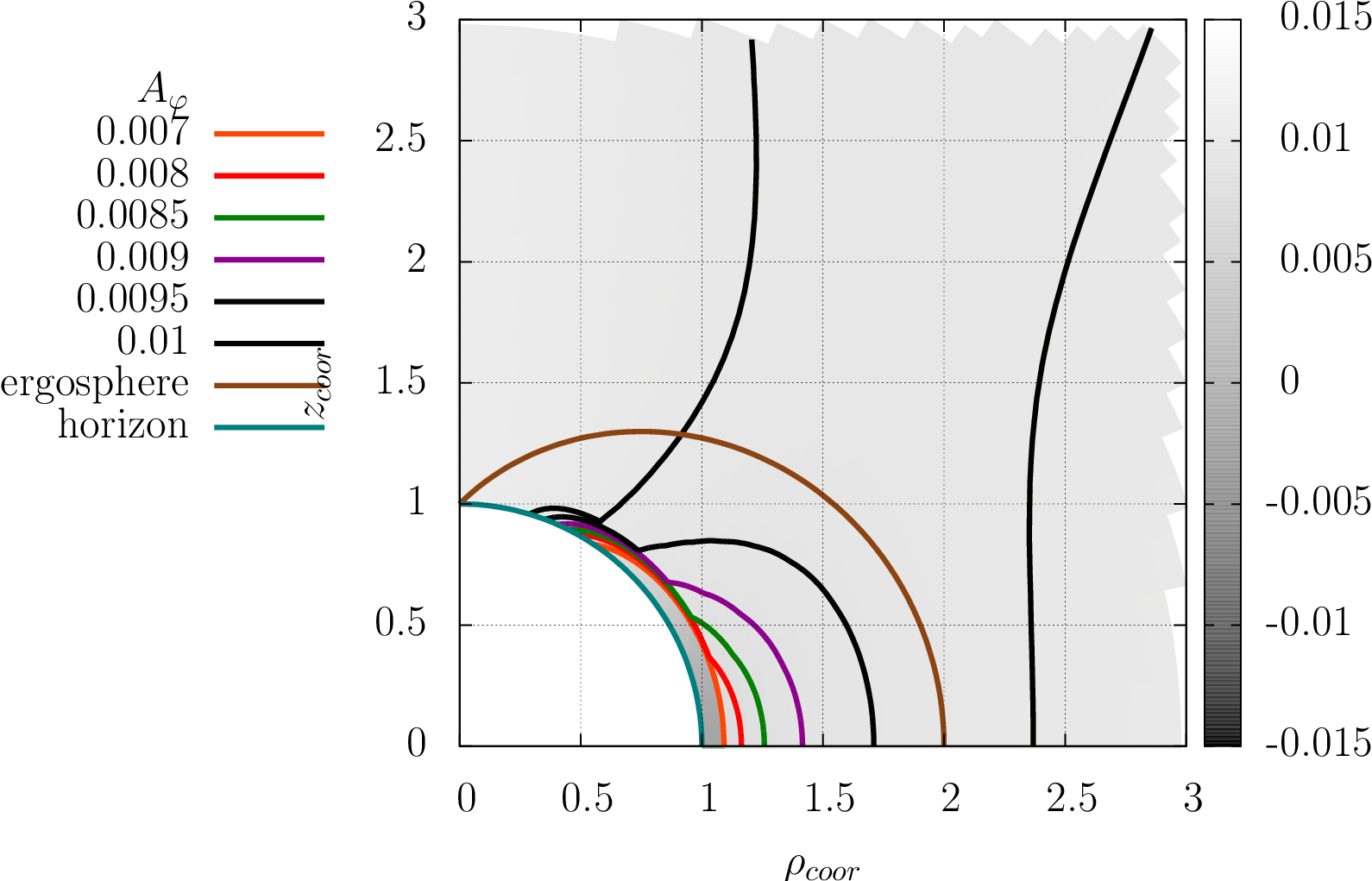}
\label{fig8a}
  }
\quad
\subfloat[]{
  \includegraphics[width=.46\textwidth]{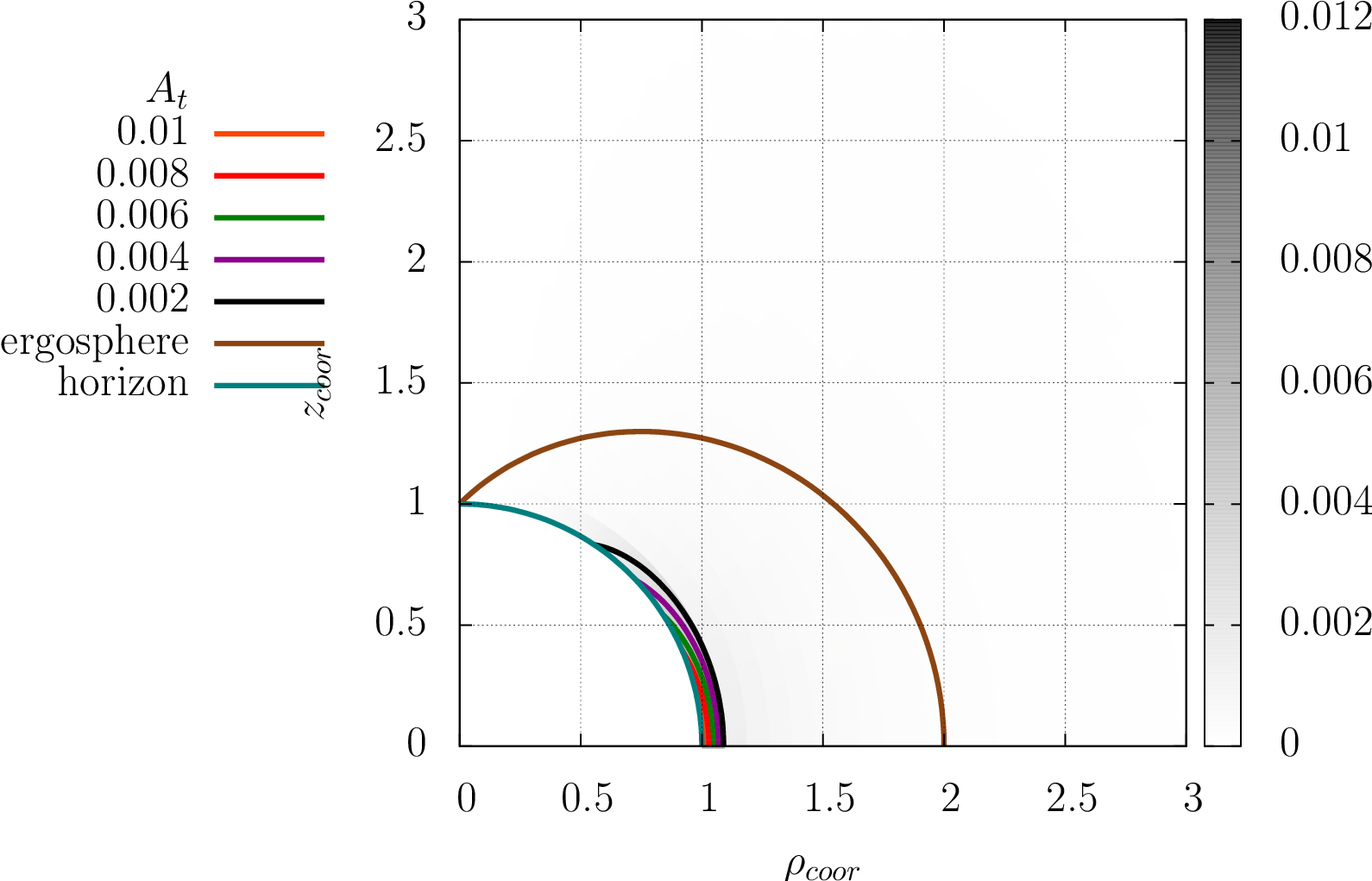}
\label{fig8b}
}\\
\centering
\subfloat[]{
  \includegraphics[width=.46\textwidth]{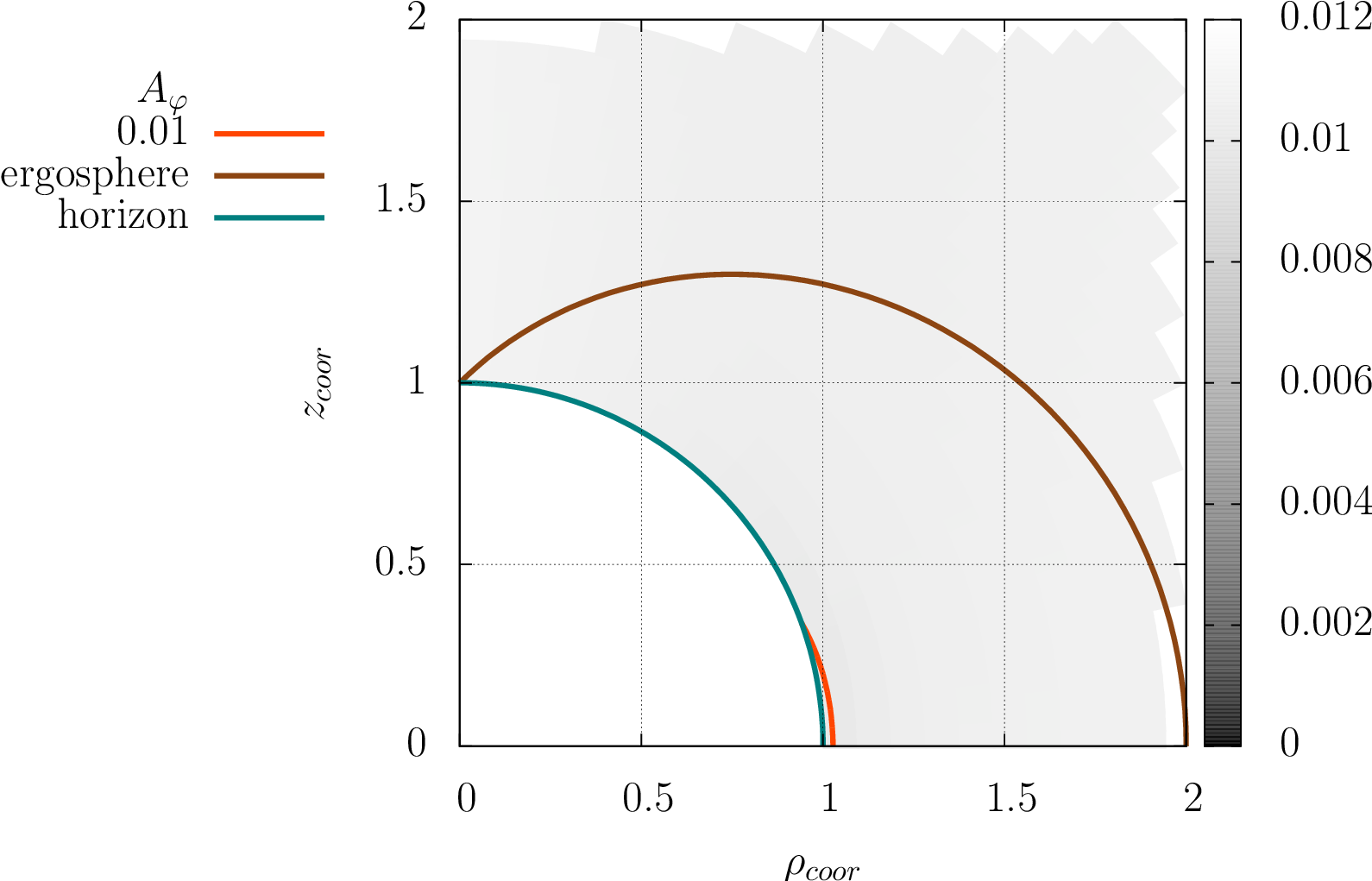}
\label{fig8c}
}
\quad
\subfloat[]{
  \includegraphics[width=.46\textwidth]{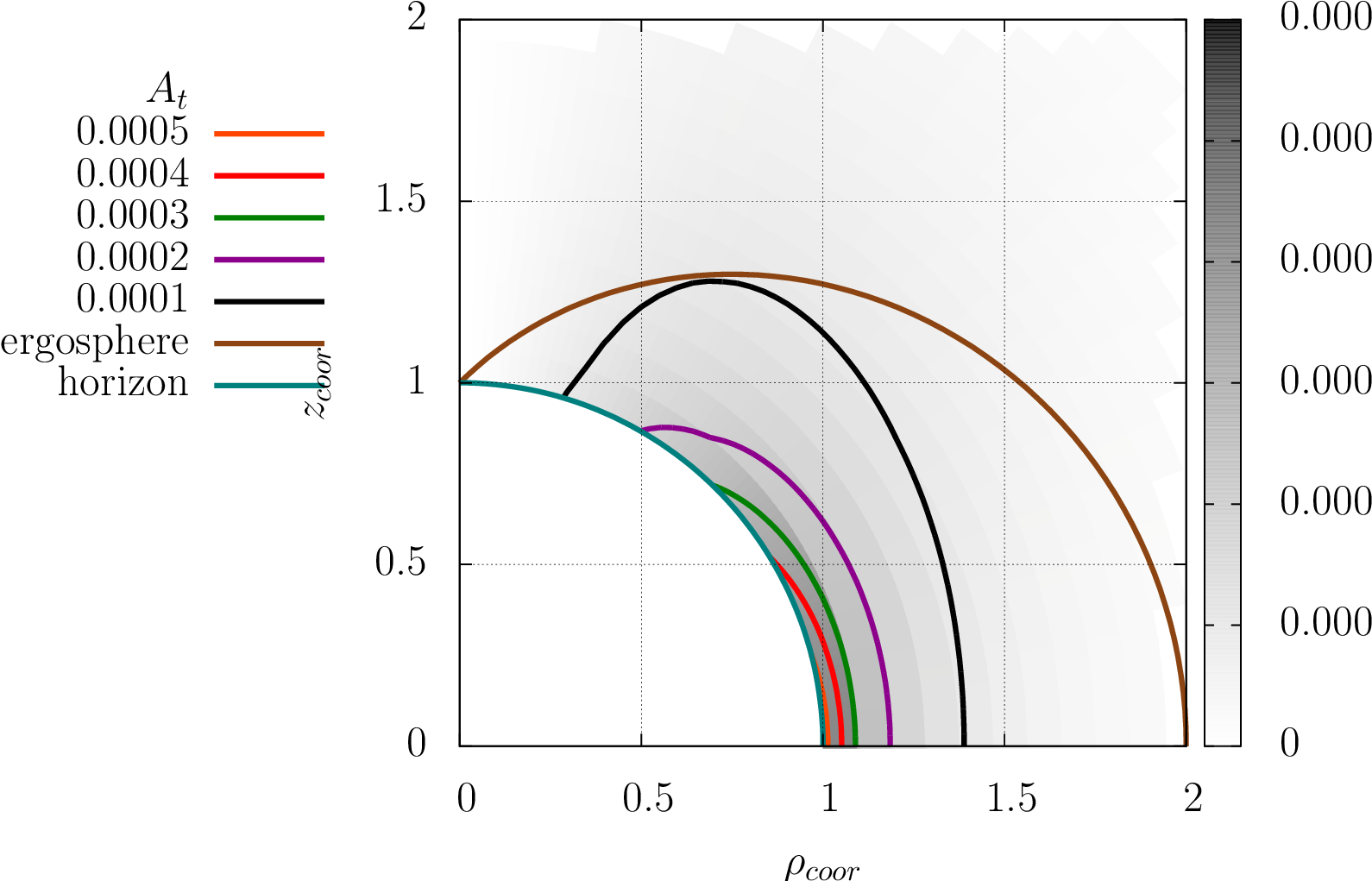}
\label{fig8d}
}
\caption{The $A_{\phi}$ and $A_{t}$ components of the electromagnetic four-potential in the extremal Kerr black hole spacetime  with $m=0.6$ (upper panels) and $m=2.1$ (lower panels).
Axes are $z_{coor} = r \cos{\theta}$, and $\rho_{coor} = r \sin{\theta}$.}
\label{fig8}
\end{figure}

Having discussed the Schwarzschild and the nonextremal Kerr black hole backgrounds, one pays attention to the extremal Kerr black hole case.
Figures \ref{fig6} and \ref{fig7} present the cosmic string and Higgs fields configurations for the black hole characterized by values of the $m$ parameter equal to $0.6$ and $2.1$. It is evident form 
figure \ref{fig6} that a small black hole expels 
the cosmic string fields, which is indicated by the line of constant $Y$ winding around the black hole horizon.
Moreover, due to the coupling of the cosmic string field to the Higgs field this expulsion is transferred to the visible sector. This implies that around a small 
extremal Kerr black hole the physical Higgs field $H_{-}$ is composed entirely of the $h$ field. 
On the other hand, for a large black hole there is no expulsion of the cosmic string, lines of constant $Y$ end on the event horizon of the 
black hole which can be seen in figure \ref{fig8}.
This dependence of the comic string expulsion on the mass of the black hole was also observed in the case of the cosmic string without coupling to the Higgs sector~\cite{gre13,nak13}. 
As for the $h$ field behavior, the formation of the wider cloud of this field around the cosmic string in the {\it dark matter sector} is also visible in figures \ref{fig6c} and \ref{fig7c}. 

In figure \ref{fig8} we presented the behavior of the $A_{\phi}$ and $A_t$ components of the electromagnetic four-potential. 
Comparing the upper and lower panels of this figure, we may see that for a small black hole the nontrivial configuration of $A_{\phi}$
stretches out far outside the ergosphere while the $A_t$ component is focused near the black hole event horizon and close to the equatorial plane. 
Meanwhile, for a large black hole mass the situation is inverted. Namely, the $A_t$ component possesses a nontrivial configuration stretching
between the event horizon and ergosphere while the $A_{\phi}$ configuration is confined to a small region near the black hole event horizon. 
Comparing figure \ref{fig8} with figures \ref{fig5}, \ref{fig4d} and \ref{fig3d}, we may draw a conclusion that this tendency is in fact also valid for
the nonextremal case. One observes that the bigger black hole mass is, the farther away $A_{t}$ reaches.
Consequently, for $A_{\phi}$ the bigger $m$ is, the smaller is the region where $A_{\phi}$ is present.

\section{Conclusions}
\label{conclusions}

In our paper we considered the problem of an Abelian-Higgs {\it dark matter} vortex in the spacetime of a stationary axisymmetric Kerr black hole.
We examined the problem both analytically and numerically. As far as the analytical approach is concerned, we mimicked {\it dark matter} by an additional $U(1)$-gauge field coupled
to the visible sector. Gauge fields in both dark and visible sectors were coupled to adequate complex scalar fields. Our main aim was to obtain a modified line element
of the Kerr black hole {\it dark matter} vortex. Using the perturbative technique we solved the underlying equations of motion, order by order in $\ep = 8\pi G \eta^2$.
In the limit of a {\it thin string}, when the mass of the Kerr black hole was far bigger than the vortex mass, we obtained the considered system metric.
The presence of the string caused a deficit not only in the angle but also in the time coordinate. The position of the black hole ergosphere was shifted.
This shift was far larger than the one caused by the Abelian-Higgs vortex, due to the fact that the mass per unit length of the {\it dark string} was larger comparing to the ordinary one.
The other interesting effect is the influence on the black hole event horizon velocity, which was bigger than in the absence of the {\it dark string}. Furthermore,
the area of the event horizon of the string threading the Kerr black hole was smaller than the area of the Kerr-Abelian-Higgs vortex system and the Kerr black hole.

We also analyzed the influence of a strong gravitational field generated by a black hole on
a cosmic string present in the {\it dark matter} sector. To obtain the field theoretical description of such an object we used the
Abelian Higgs model. Following the nomenclature of particle physics we identified its field content in the following way: 
the additional $U(1)$-gauge field represented a heavy {\it dark photon} and a complex scalar may have represented 
either a dark matter candidate or a mediation to the visible sector of the Standard Model of particle physics. 
Moreover, in our analysis we took into account a part of the Standard Model that directly couples to the {\it dark sector}.
In the case at hand it was an electromagnetic field coupling to the {\it dark photon} via the kinetic mixing term and the real component 
of the Higgs doublet (real part of the Higgs field as given in the unitary gauge) coupled to an additional complex scalar 
via a quartic coupling.

For the purpose of our investigations we numerically solved the equations of motion~(\ref{eomX})-(\ref{eomC}) and analyzed the results depicted in figures \ref{fig1} to \ref{fig8}. We found out that due to the quartic coupling among the real part of the complex scalar constituting the cosmic string ($Y$) and the real component of the Higgs field ($h$), the spatial variation in the former induces a spatial gradient 
in the latter. We may say that the structure of the cosmic string was imprinted in the Higgs field. Moreover, we may conclude that 
in all the analyzed cases the region in which the gradient in the $h$ field persisted was bigger than the analogous region for the $Y$ field.
This is in agreement with the earlier results for the flat space case obtained in~\cite{hyd14}. This behavior may have interesting 
consequences from the particle physics perspective. Namely, in the described system the masses of the $W^{\pm}$ and $Z$ bosons are 
connected to the vacuum expectation value of the $h$ field since this is the field that enters the SM Higgs doublet. On the other 
hand, the mass of the Higgs particle recently measured by LHC should be attributed to the eigenstate of the mass matrix which is
a mixture of the $h$ and $Y$ fields. This may be interpreted in the following manner. The mass of the Higgs particle is the 
same in the whole spacetime, yet the presence of the cosmic string induces a spatial variation in the vev of the $h$ field 
and this translates to the variation of the mass of the SM particles in the vicinity of the cosmic string. 
For the particular values of the coupling constants and scalar fields masses discussed in this article the gauge eigenstate Higgs field $h$ increases its value by two orders of magnitude being the biggest at the string core where $Y=0$. 
Moreover, looking at the ratio of the vacuum expectation values of the $h$ field in the string core and far away from the string we obtain 
$\frac{h_{core}}{h_{no string}} = \sqrt{ \frac{c^2 \left ( \lambda_h \tilde{\lambda}_d - 4 \lambda_{hY}^2 \right )}{\lambda_h \left ( c^2 \tilde{\lambda}_d - 2 \lambda_{hY} \right )} }$ which implies that this ratio is independent of the black hole mass.

As far as the gauge sector of the model at hand is concerned, the presence of the kinetic mixing between the {\it dark photon} and the ordinary 
electromagnetic field also led to some interesting consequences. As our gravitational background we assumed the electrically 
neutral black hole (rotating or nonrotating) which implied no electromagnetic field.
Meanwhile, the presence of the nontrivial gauge field structure of the cosmic string induced a nonvanishing electromagnetic field
outside the black hole event horizon. The field induced was directly proportional to the kinetic mixing coupling $\kappa$. 
Although the induced field is small, due to the smallness of the experimentally allowed values of $\kappa$, and focused near the black hole we may say that the presence of the cosmic string lightened up the black hole. 
From (\ref{EM_potencjal}) and (\ref{EM_potencjal2}) we may also see that the induced electric and magnetic components of the ordinary electromagnetic field will scale with the mass of the black hole as the $g_{t \phi}$ and $g_{\phi \phi}$ components of the metric.

Aside from the effect mentioned above we also observed, typical for the cosmic strings described by the Abelian Higgs model, 
a phenomenon of the cosmic string expulsion by a small extremal black hole. This implies that the cosmic string was unable to penetrate the event horizon of the small extremal black hole and ended up winding around it.

\section*{Acknowledgements}
{\L}N was supported by the National Science Centre, Poland under a postdoctoral scholarship
DEC-2014/12/S/ST2/00332. AN was supported by the National Science Centre, Poland under
a postdoctoral scholarship DEC-2016/20/S/ST2/00368. MR was partially supported by the
grant of the National Science Centre, Poland DEC-2014/15/B/ST2/00089.











\end{document}